\documentclass{article}
\usepackage{graphicx}
\usepackage{bbm}
\usepackage{amsmath}
\usepackage{amssymb}
\usepackage{amsfonts}
\usepackage{amsthm}
\usepackage{color}
\usepackage{cases}
\usepackage{setspace}
\usepackage{fullpage}

\def\theequation{{\arabic{section}}.{\arabic{equation}}}

\def\cD{\mathcal{D}}
\def\cM{\mathcal{M}}
\def\cDb{\cD_{\text {\tiny B}}}
\def\Db{D_{r {\text {\tiny B}}}}

\usepackage{hyperref}

\usepackage{subcaption}

\usepackage{algorithm}
\usepackage{algorithmic}

\newcommand{\nc}{\newcommand}
\nc{\dsp}{\displaystyle}
\nc{\R}{\Bbb{R}}
\nc{\Z}{\Bbb{Z}}
\nc{\Pp}{\Bbb{P}}
\nc{\Ap}{\Bbb{A}}
\nc{\Wp}{\Bbb{W}}
 \nc{\brho}{\boldsymbol \rho}
\nc{\va}{\vec{\boldsymbol \mu}}
\nc{\ve}{\vec{\boldsymbol \epsilon}}
\nc{\bk}{{\bf k}}
\nc{\vrho}{\vec{\brho}} 
\nc{\vr}{\vec{\bf r}}
\nc{\bx}{{\bf x}}
\nc{\vx}{\vec{\bf x}}
\nc{\om}{\omega}
\nc{\brhoi}{\brho^{\cal I}}
\nc{\bzetai}{\bzeta^{\cal I}}
\nc{\vzeta}{\vec{\bzeta}}
\nc{\vzetai}{\vec{\bzeta}^{\cal I}}
\nc{\cI}{{\cal I}}
\nc{\cJ}{{\cal J}}
\nc{\cF}{{\cal F}}
\nc{\cW}{{\cal W}}
\nc{\cA}{{\cal A}}
\nc{\cL}{{\cal L}}
\nc{\cS}{{\cal S}}
\nc{\cC}{{\cal C}}
\nc{\cN}{{\cal N}}
\nc{\vrhoi}{\vec{\brho}^{\,\cal I}}
\nc{\xii}{\xi^{\cI}}
\nc{\etai}{\eta^{\cI}}
\nc{\la}{\lambda}
\nc{\de}{\delta}
\nc{\ep}{\varepsilon}
\nc{\vu}{\vec{\bf u}}
\nc{\bu}{{\bf u}}
\nc{\vui}{\vec{\bf u}^{\cI}}
\nc{\bui}{{\bf u}^{\cI}}
\nc{\bt}{{\bf t}}
\nc{\vt}{\vec{\bt}}
\nc{\bn}{{\bf n}}
\nc{\vn}{\vec{\bn}}
\nc{\bm}{{\bf m}}
\nc{\vm}{\vec{\bm}}
\nc{\vrp}{\vec{{\bf r}'}}
\nc{\vrc}{\vec{{\bf r}^c_p}}
\nc{\ts}{\tilde s}
\nc{\os}{\overline s}
\nc{\tom}{\tilde \om}
\nc{\tO}{\tilde \Omega}
\nc{\tS}{\tilde S}
\nc{\oS}{\overline S}
\nc{\vrhos}{\vrho_{\star}}
\nc{\vrhosi}{\vrho_{\star}^{\cI}}
\nc{\brhos}{\brho_{\star}}
\nc{\brhosi}{\brhos^{\cI}}
\nc{\vms}{\vm_{\star}}
\nc{\vmi}{\vm_{_{\cI}}}
\nc{\vM}{\vec{\bf M}}
\nc{\vMi}{\vM_{_{\cI}}}
\nc{\Ppi}{\Pp_{\cI}}
\nc{\vxi}{\vec{\bxi}}
\nc{\bK}{{\bf K}}
\nc{\bmi}{\bm_{_{\cI}}}
\nc{\Pppi}{\Ppi^p}
\nc{\be}{{\bf e}}
\nc{\bep}{{\bf e}^p}
\renewcommand{\hat}{\widehat}

\begin{document}
\title{Low rank plus sparse decomposition of synthetic aperture radar data for target imaging and tracking}
\author{Matan Leibovich\footnotemark[1], George Papanicolaou\footnotemark[2], and Chrysoula Tsogka\footnotemark[3]
}\renewcommand{\thefootnote}{\fnsymbol{footnote}}
\footnotetext[2]{Institute for Computational and Mathematical Engineering, Stanford University, Stanford, CA 94305. \\\   (matanle@stanford.edu)} 
\footnotetext[2]{Department of
	Mathematics, Stanford University, Stanford, CA 94305.
	(papanicolalou@stanford.edu)}\footnotetext[3]{Applied Math Unit,
	University of California, Merced, 5200 North Lake Road, Merced, CA
	95343 (ctsogka@ucmerced.edu)}\date{}
\maketitle  \date{}

	
	\begin{abstract}
		We analyze synthetic aperture radar (SAR) imaging of complex ground scenes that contain both stationary and moving targets. In the usual SAR acquisition scheme, we consider ways to preprocess the data so as to separate the contributions of the moving targets from those due to stationary background reflectors. Both components of the data, that is, reflections from stationary and moving targets, are considered as signal that is needed for target imaging and tracking, respectively. The approach we use is to decompose the data matrix into a low rank and a sparse part. This decomposition enables us to capture the reflections from moving targets into the sparse part and those from stationary targets into the low rank part of the data. The computational tool for this is robust principal component analysis (RPCA) applied to the SAR data matrix. We also introduce a lossless baseband transformation of the data, which simplifies the analysis and improves the performance of the RPCA algorithm. Our main contribution is a theoretical analysis that determines an optimal choice of parameters for the RPCA algorithm so as to have an effective and stable separation of SAR data coming from moving and stationary targets. This analysis gives also a lower bound for detectable target velocities. We show in particular that the rank of the sparse matrix is proportional to the square root of the target's speed in the direction that connects the SAR platform trajectory to the imaging region. The robustness of the approach is illustrated with numerical simulations in the X-band SAR regime. 
	\end{abstract}

	\section{Introduction}
	\subsection{Synthetic aperture radar imaging}
\begin{figure}[htbp]
		\begin{center}
			\resizebox{0.36\textwidth}{0.32\textwidth}{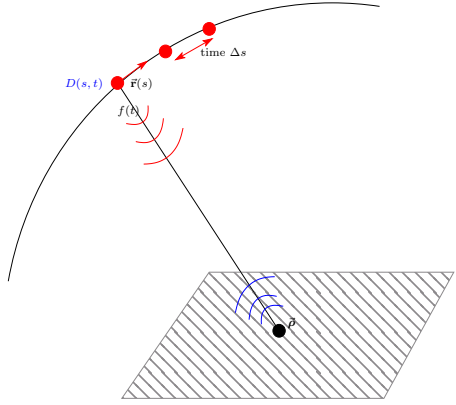}
		\end{center}
		\caption{Synthetic aperture imaging schematic.}
		\label{fig:sar}
	\end{figure}

	Synthetic aperture radar can provide high resolution imaging in several applications~\cite{cheney2009fundamentals}. 
	The SAR data are usually acquired by a moving platform that probes a remote area by sending broadband pulses, $f(t)$, and recording the echoes. Although there is only one transmit-receive element, high resolution images are obtained by coherently processing the data collected by the moving platform along a large synthetic aperture. Let $\vr(s)$ denote the platform location at time $s$, called the slow time.  The data collected form a matrix $D(s,t)$, which records the echoes received at $\vr(s)$ when a pulse is emitted from the same location. We consider $n+1$ discrete slow times with sampling rate $\Delta s$, 
	$s_{j} = j \Delta s$, with $j = -n/2, \cdots, n/2$. The fast time, $t \in (0,\Delta s)$ runs between successive pulse emissions. A schematic of a SAR imaging configuration is shown in Figure~\ref{fig:sar}.
	
	To maximize the power delivered to the region to be imaged, the probing signals
	used in SAR are long pulses of support $t_{c} \gg 1/B$ where $B$
	denotes the bandwidth of the probing pulse. An example of such signals are
	linear frequency modulated chirps~\cite{cheney2009fundamentals}. To 
	concentrate the energy of the reflected echoes to a small time interval of size $1/B$ 
	the received signal is convolved with the complex conjugate of the time-reversed 
	emitted pulse. 
	This step is called pulse compression and is followed 
	by a range compression step which consists of migrating the pulse compressed data 
	to a reference point $\vrho_o$ in the imaging window. The range-compression step 
	removes from the data the large phase $\omega \tau(s,\vrho_o)$.  
	These two steps together are referred to as down-ramping, 
	\begin{equation}
	\begin{split}
	D_r(s,t) &= \int \mathrm{d}t'
	D\left(s,t-t'+\tau(s,\vrho_o)\right)
	\overline{f(-t')}\\
	& = \int \frac{\mathrm{d}\omega}{2 \pi} \overline{
		\widehat{f}(\omega)} \widehat{D}(s,\omega) e^{-\mathrm{i} \omega
		[t+\tau(s,\vrho_o)]},
	\end{split}
	\label{eq:down-ramping}
	\end{equation}
	with $\widehat f (\omega)$ (res. $\widehat{D}(s,\omega)$) denoting the Fourier transform of $f(t)$ (resp. $D(s,t)$), defined as 
	\begin{equation}
	\begin{split}
	\hat f (\om) &= \int d t \, f(t) e^{i \om t}.
	\end{split}
	\end{equation}
	Here, $\tau(s,\vrho_o)$ is the round trip travel time between the platform location at slow time $s$, $\vr(s)$ and
	the reference point location $\vrho_o$,   
	\begin{equation}
	\dsp \tau(s,\vrho_o) = 2 \frac{| \vr(s) - \vrho_o |}{c},
	\label{eq:tau}
	\end{equation}
	with $c$ the speed of light. 
	
	The SAR image is formed by summing coherently the down-ramped data $D_r(s_j,t)$ 
	back-propagated to the imaging point $\vrho$ using the travel times differences 
	$\tau(s_j,\vrho)-\tau(s_j,\vrho_o)$,
	\begin{equation}
	\dsp  I^{\text SAR}(\vrho) = \sum_{j=-n/2}^{n/2} 
	D_r(s_j,\tau(s_j,\vrho)-\tau(s_j,\vrho_o)) .
	\label{eq:SARfunctional}
	\end{equation}
	The usual SAR image function \eqref{eq:SARfunctional} assumes that only reflections from stationary targets are contained in the down-ramped data $D_r(s,t)$. Consequently, if moving targets are present in the region to be imaged then their reflections are not correctly back-propagated and this results in images with streaks or ghosts that depend on the reflectivity and the velocity of the moving targets. For a complex scene, with many stationary and moving targets the overall image is severely distorted and neither the stationary nor the moving targets can be imaged or tracked accurately. To address this problem several motion estimation and separation strategies have been developed. Before considering them we describe first the structure of the SAR data matrix.

	
		\subsection{The SAR data matrix}
	We consider a model for the SAR data matrix for a scene comprised of $N$ small point-like targets. We denote $\sigma_i$ the reflectivity of the  $i$th target and $\vrho_i(s)$ its location at slow time $s$.  The SAR platform emits a pulse $f(t)$ and the down-ramped data, as in \eqref{eq:down-ramping}, are obtained by convolving the received echoes with $\overline{f(-t)}$. It is therefore as if the antenna emitted the pulse $f_p(t)$ defined as
	\begin{equation}
	f_p(t) = \int dt' f(t') \overline{f(t'-t)} .
	\label{eq:f_p}
	\end{equation}
	The down-ramped SAR data matrix can be modeled as a superposition of pulses $f_p$ shifted by $\Delta \tau_i(s)$ and multiplied by $\sigma_i$, the reflectivity of each target 
	\begin{equation} \label{eq:MOD1}
	D_r(s,t)=\sum\limits_{i=1}^N\sigma_i  f_p(t - \Delta \tau_i(s)). 
	\end{equation}
	The difference travel time $\Delta \tau_i(s)$ is the round trip travel time from the antenna location $\vr(s)$ to the target $\vrho_i(s)$ from which the round trip travel time from the antenna location $\vr(s)$ to the reference location $\vrho_o$ is subtracted to account for the range compression step of down-ramping  
	\begin{equation}
	\Delta \tau_i(s)=\frac{2(\|\vr(s)-\vrho_i(s)\|-\|\vr(s)-\vrho_o\|)}{c}.
	\label{eq:Delta_tau}
	\end{equation}
	We used here the start-stop approximation which neglects the targets' displacement during the round trip travel time. This is justified in radar because the electromagnetic waves 
	travel at the speed of light that is many orders of magnitude larger than the speed of the targets and the platform.
	
	It is usually assumed that $f_p(t)$ consists of a base-band waveform
	$f_B(t)$ modulated by a carrier frequency $\nu_o = \om_o/(2 \pi)$,
	\begin{equation}
	f_p(t) =  \cos(\om_o t) f_B(t).
	\label{eq:1.1}
	\end{equation}
	Its Fourier transform is
	\begin{equation}
	\begin{split}
	\hat f_p(\om) &= \int d t \, f_p(t) e^{i \om t}
\\&	= \frac{1}{2} \left[\hat f_B(\om + \om_o) +
	\hat f_B(\om - \om_o)\right].
	\end{split}
	\end{equation}
	Here $\hat f_B(\om)$ is supported in the interval $[-\pi B,\pi B]$, where
	$B$ is the bandwidth while  $\hat f(\om)$ is supported in $[-\omega_o-\pi B,-\omega_o+\pi B] \cup [\omega_o-\pi B,\omega_o+\pi B]$.
	
	The down-ramped SAR data matrix takes the form
	\begin{equation}
	D_r(s,t)=\sum\limits_{i=1}^N\sigma_i \cos(\omega_o(t-\Delta \tau_i(s)))f_B(t-\Delta \tau_i(s)),
	\label{eq:Dr_time}
	\end{equation}
	and its Fourier transform is 
	\begin{equation}
	\hat{D}_r(s,\omega)=\sum\limits_{i=1}^N\frac{e^{-i\omega\Delta \tau_i(s)}}{2}\left[\hat{f}_B(\omega+\omega_o)+\hat{f}_B(\omega-\omega_o)\right].
	\label{eq:Dr_freq}
	\end{equation}

	Our model for the SAR data matrix $\cD \in {\mathbb{R}}^{(n+1)\times(m+1)}$ is defined by the discrete samples of (\ref{eq:Dr_time}), 
	\begin{equation}
   \begin{split}
	&\cD_{il} =D_r\left(s_{i-\frac{n}{2}-1},
	t_{l-1}\right), 
	\\ &i = 1, \ldots, n+1, ~ ~ l = 1, 
	\ldots, m + 1,
	\end{split}
	\label{eq:MOD6}
	\end{equation}
	with slow times $s_j$ defined by
	\begin{equation}
	s_j = j \Delta s,\quad j= -n/2,\ldots, n/2, 
	\end{equation}
	and fast times $t_l$ defined as
	\begin{equation}
	t_l = l \Delta t,\quad l=1, \ldots, m. 
	\end{equation}
	Here we assumed that the pulse repetition rate $\Delta s$ is an integer multiple of $\Delta t$ and set $m= \Delta s/ \Delta t$.

	\subsection{Motion estimation and separation strategies}
	A simple way to separate moving targets from the stationary background is the change detection (CD) method. In video surveillance, this amounts to subtracting the data corresponding to two consecutive frames.  Assuming that multi-frame SAR images, which show the same scene at different times,  are available, CD can be used in SAR. We refer to \cite{Jahangir2009} where CD is used for detecting moving targets in multi-frame SAR imagery and to \cite{Amin2013} where CD was implemented for human motion tracking with through-the-wall radar. A similar method is the Coherent Change Detection (CCD) which also involves performing repeat pass radar data collection. However, the change detection part is now different. Change is detected in CCD by  identifying zeros in the cross-correlation between pairs of images \cite{Preiss, Damini2013, Cha2014}.    
	
	A specific form of CD that has been used extensively in SAR is the Displaced Phase Center Antenna (DPCA) method, in which the platform uses two antennas, synchronized so that the second antenna is following the same trajectory as the first with a certain time delay. Subtracting the two traces eliminates the stationary background echoes, leaving only the echoes from moving targets \cite{muehe2000displaced,cheney2009fundamentals}
	
	Another method that has been used  for moving target detection is autofocus \cite{fienup, jao, kirscht, li2007}. These algorithms exploit the  fact that target motion introduces phase errors, which result in smearing of the target's image. Autofocus algorithms can be used in the classical single channel SAR setup, and they are composed by two steps: first the conventional smeared images are formed and subsequently a phase error estimation and compensation of the image parts that correspond to moving targets is performed.

	A lot of progress has been also achieved in space-time adaptive processing algorithms which aim to better detect the moving targets by suppressing the clutter of stationary targets. Examples are the multi-channel SAR \cite{ender1993}, the space-time-frequency SAR \cite{ender1996}, velocity SAR (VSAR) \cite{friedlander} and dual-speed SAR \cite{wang2006}.  More recently compressed sensing and sparsity driven methods made their way into the moving target SAR imaging problem \cite{Onhon2011}. We refer the reader to \cite{Willsky2014} for a review on sparsity driven synthetic aperture imaging including moving target imaging.


	
	
	
	\subsection{Robust principal component analysis}
	We follow here the approach proposed in \cite{borcea2013synthetic}, where the robust principal component analysis was used for separating the data matrix $\cD$ in \eqref{eq:MOD6} in two subsets: the echoes due to stationary targets that form the low rank part of the data matrix and those due to moving targets echoes which constitute the sparse part. The RPCA method consists of solving the following convex optimization problem,
	
	\begin{eqnarray}
	& \min_{\cL,\cS\in\mathbb R^{n_1\times n_2}}
	\quad ||\cL||_*+ \eta ||\cS||_1\label{pca1} \\
	& \text{subject to} \quad \cL+\cS=\cD. \label{pca2}
	\end{eqnarray}
	Here $||\cL||_*$ denotes the nuclear norm, that is the sum of the singular values of $\cL$, and $||\cS||_1$ is the matrix $\ell_1$-norm of $\cS$.  Assuming the matrix $\cD \in \mathbb{R}^{n_1 \times n_2}$ is the sum of a low rank matrix $\cL_o$ and a sparse matrix $\cS_o$ then the solution of the optimization \eqref{pca1}-\eqref{pca2} recovers $\cL_o$ and $\cS_o$ exactly, under sufficient conditions. This optimization problem has been analyzed in \cite{candesRPCA}, where the sufficient conditions for the decomposition to be exact are derived. 
	
	The parameter $\eta$ that balances the ratio between the nuclear norm of $\cL$ and the 1-norm of $\cS$ will play an important role in our analysis. The main idea being that we would like to adequately chose the value of $\eta$ given that $\cL$  and $\cS$ represent the echoes received from stationary and moving targets. The recommended value for $\eta$ proposed in \cite{candesRPCA} and used for SAR data in \cite{borcea2013synthetic} is  
	\begin{equation}
	\eta = \frac{1}{\sqrt{\max
			\{n_1, n_2\}}}.
	\label{eq:eta}
	\end{equation}
	It was shown in \cite{borcea2013synthetic} that, with this choice of $\eta$, the RPCA algorithm is sensitive to the window size of the data. Moreover, for complex scenes successful separation between moving and stationary targets' echoes can be achieved only with adequate windowing of the SAR data. 
	
	\subsection{Summary of main results}

We determine in this paper an optimal range for the parameter $\eta,$ by taking into consideration the specific form that the $\cL_o$ and $\cS_o$  matrices take in the SAR problem.  Specifically, by analytically computing the nuclear and $\ell_1$ norms for a single stationary and moving target, we obtain a range of values for $\eta$ for which robust separation results can be obtained with RPCA.
	Furthermore, our analysis shows that the range of acceptable values for $\eta$ increases with increasing velocity of the moving target. This is intuitively expected since for increasing target velocity the echoes from the moving target become {\em more sparse} or equivalently {\em less low rank}. Thus, they can be more easily separated from the stationary targets' echoes. This implies in particular that targets that are moving with a speed above a certain threshold can be easily detected. More precisely, it is the projection of the velocity along the direction that connects the SAR platform to the imaging domain that matters. This is because for targets that are moving parallel to that direction, smaller variation in travel times is expected.   
	
	Our analysis suggests that the RPCA algorithm should be applied on the baseband data matrix $\cDb$ which is obtained from the down-ramped data, $\cD$, using a lossless transformation that consists of shifting the signal in the frequency domain so that it is centered at $\omega=0$ instead of $\omega=\omega_o$. 
	Going to baseband allows us to evaluate the performance of the RPCA algorithm for a wider range of moving target speeds. This transformation also improves the accuracy of our theoretical analysis and our numerical results indicate that it also improves the robustness of RPCA.
	
	We derive a closed form expression for an optimal value of the RPCA parameter, $\eta^*$, for both the baseband and the original SAR data matrix. We show that using $\eta^*$ does yield results that are more robust, especially for the baseband case. Robustness is observed with respect to the window size (aperture) of the SAR data, the complexity of the scene and the moving target's velocity.

\section{Robust PCA for motion detection and SAR data separation}
\label{sect:decomp}
 We present here the robust PCA algorithm adapted to the SAR problem. We first explain the baseband transformation. 
The performance of RPCA on the baseband data matrix is  theoretically equivalent  to the one on the original data matrix, but going to baseband greatly simplifies the analysis. We also see numerically that it yields more robust behavior, in a sense that will be explained in the analysis section \ref{RPCA:anal}.  
\subsection{Baseband data matrix}
\label{sec:BB_matrix}
The proposed preprocessing takes the original data matrix and creates a baseband (BB) version of it by performing band pass filtering at every slow time $s$. 

Assuming the model of \eqref{eq:MOD1} for a single scatterer, the Fourier transform of the down ramped data is 
$$\hat{D}_r(s,\omega)=\frac{e^{-i\omega\Delta \tau(s)}}{2}\left[\hat{f}_B(\omega+\omega_o)+\hat{f}_B(\omega-\omega_o)\right].
$$
Thus, the Fourier transform of $D_o(s,t)=e^{i\omega_o t}D_r(s,t)$ is
$$
\hat{D}_o(s,\omega)=\frac{e^{-i(\omega-\omega_o) \Delta\tau(s)}}{2}\left[\hat{f}_B(\omega)+\hat{f}_B(\omega-2\omega_o)\right].
$$
We then filter out the signal centered around $2\omega_o$ to get a baseband signal centered around $\omega=0$, 
$$
\hat{D}_{r {\text {\tiny B}}}(s,\omega)=\hat{h}_{LP}(\omega)\hat{D}_o(s,\omega)= e^{-i(\omega-\omega_o) \Delta\tau(s)}\hat{f}_B(\omega).
$$
There is no intersection between the supports since we assume $\omega_o\gg B$ (narrow band regime). Thus,  the filter can be e.g., 
$$
\hat{h}_{LP}(\omega)=\begin{cases} 2& |\omega|\le 2B\\ 0& |\omega|>\alpha B\end{cases}
,$$
where $\alpha B<2\omega_o-B$ is small enough to eliminate the signal centered around $2\omega_o$. The factor of 2 ensures the maximum absolute value remains 1.

In the time domain we will get a filtered signal of the form
\begin{equation}
D_{r{\text {\tiny B}}}(s,t)=e^{i\omega_o\Delta\tau(s)}f_B(t-\Delta \tau(s)).
\label{eq:BB_form}
\end{equation}
No data is lost in the process, and the original data matrix can be reconstructed by taking the real part of the BB data multiplied by $e^{-i\omega_o t}$ ,
$$
D_r(s,t)=\Re \left\{ e^{-i\omega_o t}D_{r{\text {\tiny B}}}(s,t)\right\}.
$$

We define $\cDb$, the baseband data matrix, by applying the Discrete Fourier Transform (DFT) to the data matrix $\cD$ with respect to its columns, and applying a similar discrete filter to go to baseband. Assuming a small enough $\Delta t$ 
the DFT is a good approximation of the Fourier transform and we have
\begin{equation}
\begin{split}
&
\left(\cDb\right)_{il}=D_{r{\text {\tiny B}}}(s_{i-\frac{n}{2}-1},t_{l-1}),\\& i = 1, \ldots, n+1, ~ ~ l = 1, 
	\ldots, m + 1.
	\end{split}
\end{equation} 
\subsection{Modification of RPCA for complex matrices}
Following  \cite{lin2010augmented} we use the inexact augmented Lagrangian implementation of RPCA  as described in Algorithm \ref{alg:inexact_alm}. 
This algorithm iteratively approximates the solution by alternatively minimizing the following augmented Lagrangian with respect to $\cL$ and $\cS$
$$\|\cL\|_*+\eta\|\cS\|_1+\langle Y, \cD-\cL -\cS \rangle + \frac{\mu}{2}\|\cD-\cL -\cS\|_F^2,
$$
where $\langle A, B\rangle =\text{tr}(A^TB)$ is the regular matrix norm.
 Each of the minimization steps has an analytical solution given by either spectral- or element-wise thresholding. 

The RPCA algorithm can be adapted for use on complex matrices by modifying the thresholding associated with $L^1$ minimization to preserve the argument of the number,
\begin{equation}
\Theta_{\eta}(a)=e^{i \arg a} \max(|a|-\eta,0).
\end{equation}

The penalty term coefficient increases with every iteration, usually geometrically
$\mu_{k+1}=\rho \mu_k$. 
\begin{algorithm}[htbp]
	\caption{$\cD=\cL+\cS$ Inexact ALM method (algorithm no.~5 in \cite{lin2010augmented}, see also for initial parameters' choice and update).}
	\begin{algorithmic}[1]
		\setstretch{1.2}
		\STATE{\textbf{Input}: Observation matrix $\cD\in \mathbb{C}^{m\times n}$}
		\STATE{$Y_0=\cD/J(\cD);\cS_o=0;\mu_0>0;\rho>1;k=0\quad J(\cD) = \max \left(
			\|\cD\|_2, \eta^{-1}\|\cD\|_\infty\right)$}
		\WHILE{not converged}
	   \STATE{$[U_{k+1},\Sigma_{k+1},V_{k+1}]=\text{svd}(\cD-\cS_{k}-\mu_k^{-1}Y_k)$}
		\STATE{$\cL_{k+1}=U_{k+1}\Theta_{\mu_k^{-1}}[\Sigma_{k+1}]V_{k+1}^*$}
		\STATE{$\cS_{k+1}=\Theta_{\eta\mu_k^{-1}}[\cD-\cL_{k+1}+\mu_k^{-1}Y_k]$}
		\STATE{$Y_{k+1}=Y_k+\mu_k(\cD-\cL_{k+1}-\cS_{k+1})$}
		\STATE{ $\mu_{k+1}=\rho \mu_k$}
		\STATE{$k\rightarrow k+1$}
		\ENDWHILE
		\RETURN $\cL_{k+1},\cS_{k+1}$
	\end{algorithmic}
	\label{alg:inexact_alm}
\end{algorithm}

Based on  \cite{lin2010augmented}, initial parameters were set to $\mu_0=1.2/\|\cD\|_2$ and $\rho=1.4$. 

\subsection{Robustness of RPCA with optimal $\eta$}
\begin{figure}[htbp]
	\centering
	\begin{subfigure}[t]{0.57\textwidth}
		\includegraphics[width=1\columnwidth]{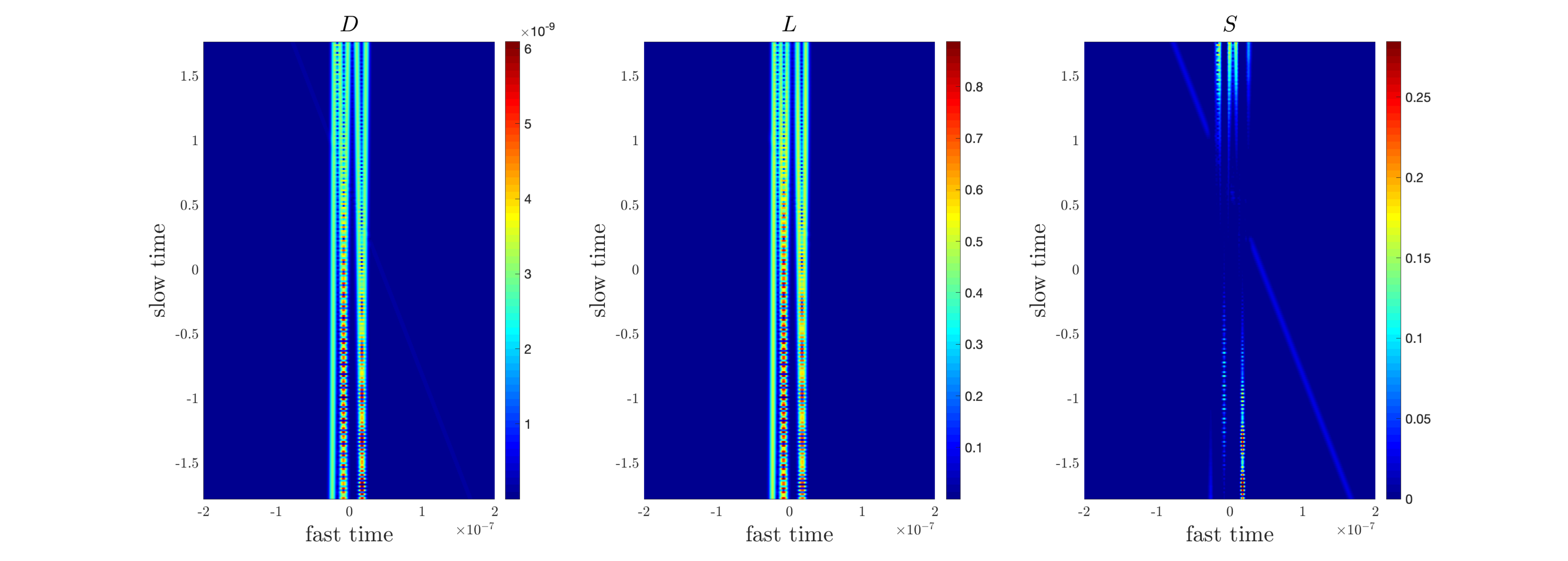}
		\caption{Original (left), low rank (middle) and sparse (right) part of the data obtained with RPCA using $\eta=\frac{1}{\sqrt{\max(n_1,n_2)}}$. RPCA applied to the entire data matrix $\cD$ (1 window).}
		\label{fig:DLS_no_window}
	\end{subfigure}
	
	\begin{subfigure}[t]{0.57\textwidth}
		\includegraphics[width=1\columnwidth]{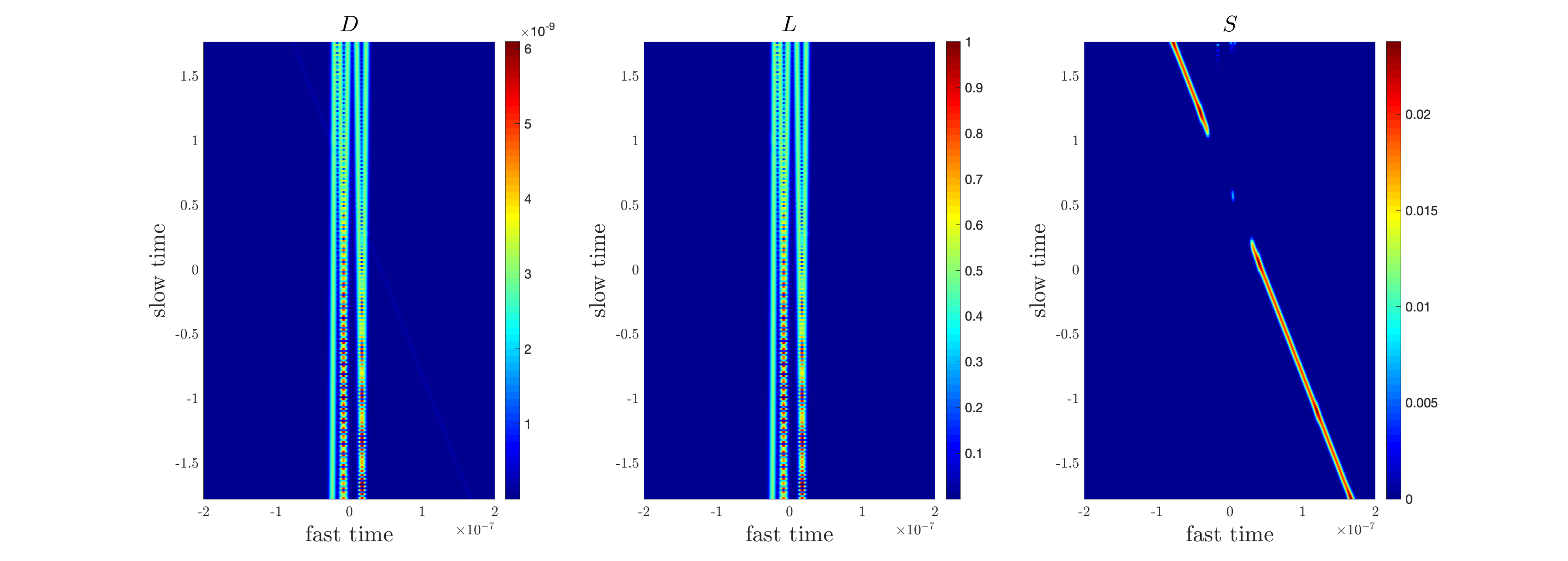}
		\caption{Same as (a) but RPCA is applied after decomposing the data into 5 consecutive windows in slow time.}
		\label{fig:DLS_4_window}
	\end{subfigure}
	
	\begin{subfigure}[t]{0.57\textwidth}
		\includegraphics[width=1\columnwidth]{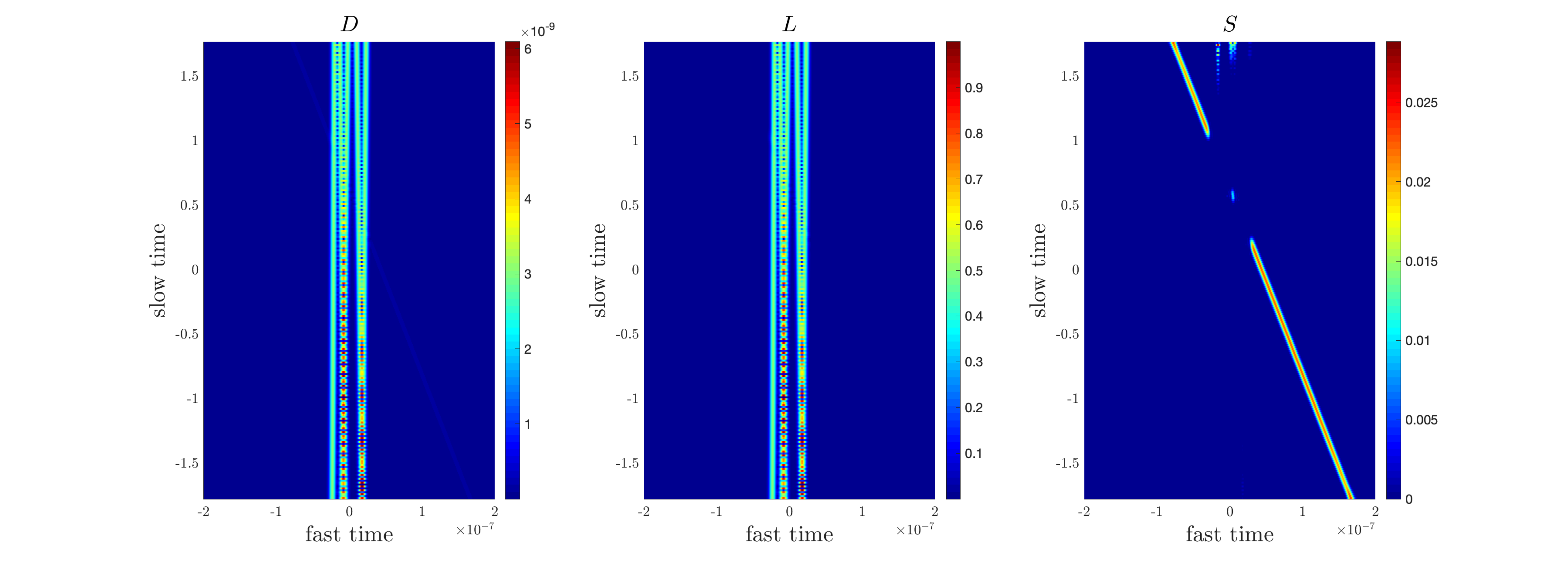}
		\caption{RPCA applied to the entire data matrix $\cD$ using the optimal $\eta^*$.}
		\label{fig:DLS_opt}
	\end{subfigure}
	\caption{Example of the performance of RPCA applied to a scene with five stationary targets and one moving target with speed $15$~m/s. The moving target's reflectivity is 5\% of the stationary targets'.
	This example illustrates that very good $\cL+\cS$ separation can be achieved using wider apertures (larger windows) when the optimal parameter $\eta$ is used.}\label{fig:RPCA}
\end{figure}

Before presenting our analysis, we wish to illustrate with a numerical example the robustness of RPCA when using the optimal parameter $\eta$. We consider a scene with five stationary targets and one moving target at a speed $15$~m/s. The moving target's reflectivity is 5\% of the stationary targets', which makes the detection of the moving target quite challenging. We show the SAR data matrix and its $\cL+\cS$ decomposition in Figure \ref{fig:RPCA}. The performance of RPCA with the conventional choice for $\eta$, i.e., $\eta=\frac{1}{\sqrt{\max(n_1,n_2)}}$ is not very good when applied to the entire data matrix (see Figure \ref{fig:DLS_no_window}) but it is significantly improved when the data are decomposed into $5$ consecutive windows in slow time as seen in Figure \ref{fig:DLS_4_window}. The plots in  Figure \ref{fig:DLS_opt} illustrate that RPCA achieves very good separation without any data windowing when the optimal parameter $\eta^*$ is used.

\begin{figure}[htbp]
	\centering
	\begin{subfigure}[t]{0.35\textwidth}
		\includegraphics[width=1\columnwidth]{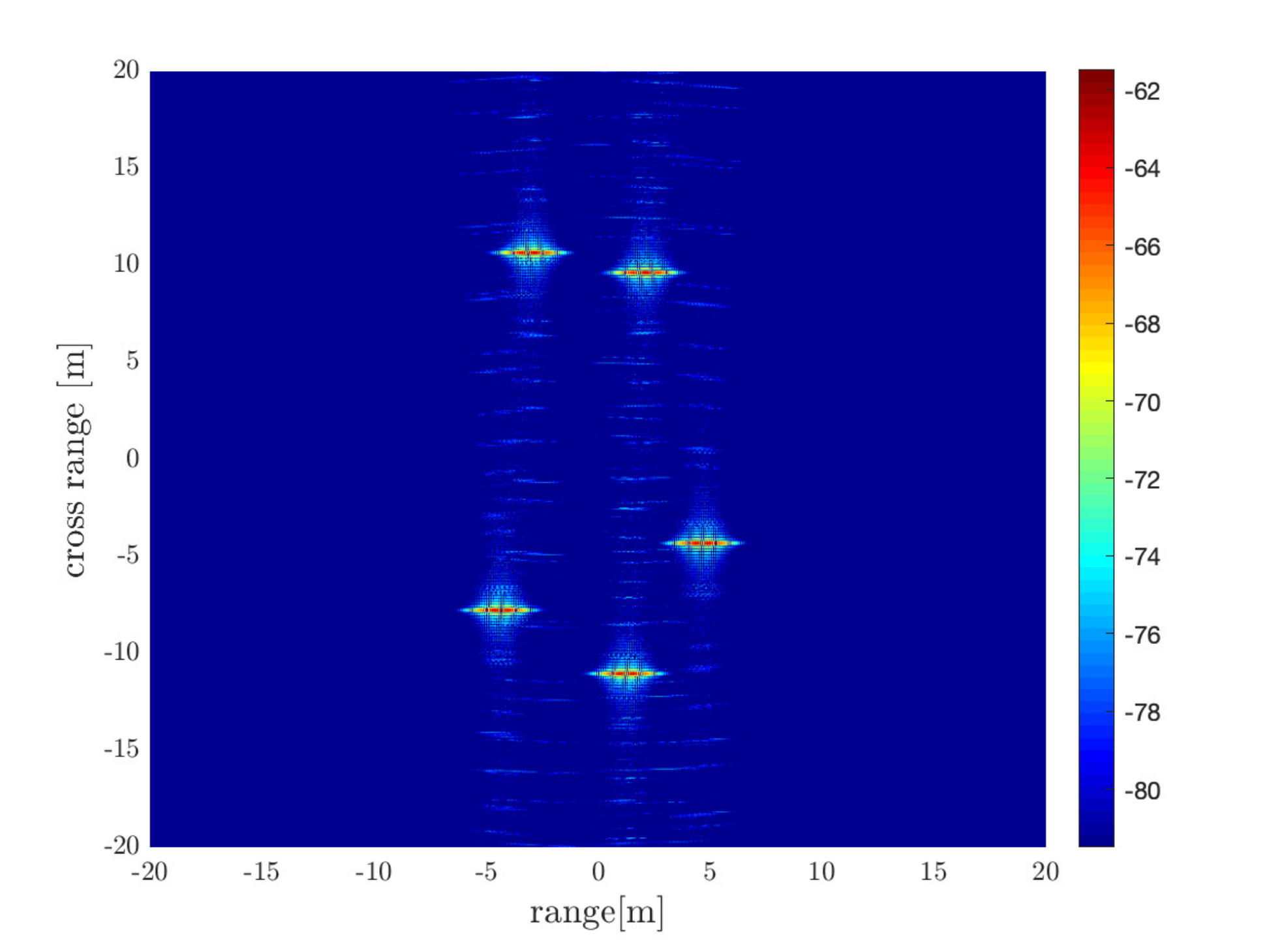}
		\caption{Kirchhoff Migration applied to the original data matrix. }
		\label{fig:image_D}
	\end{subfigure}
	\begin{subfigure}[t]{0.35\textwidth}
		\includegraphics[width=1\columnwidth]{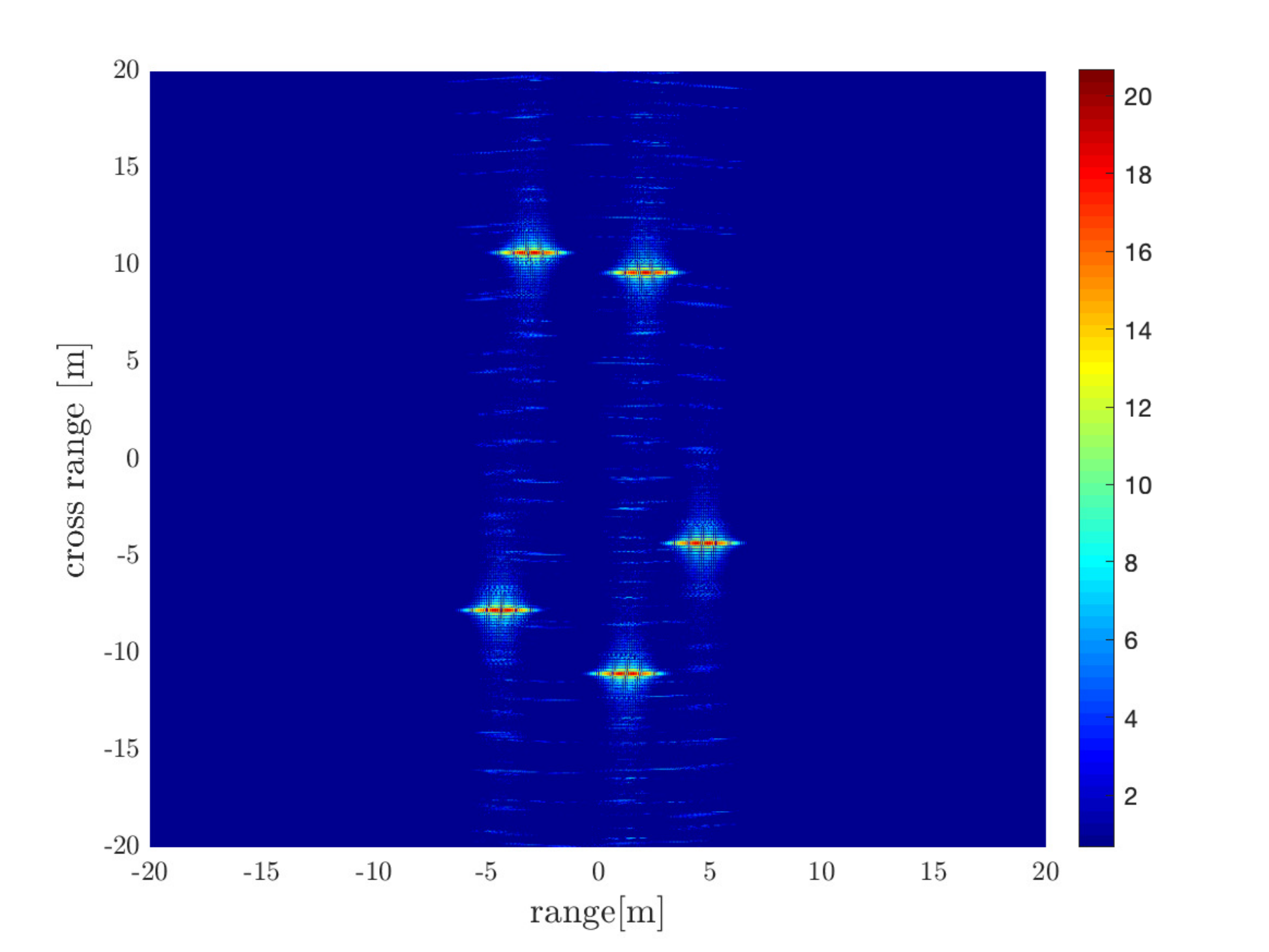}
		\caption{KM applied to the low rank part.}
		\label{fig:image_L}
	\end{subfigure}
	\begin{subfigure}[t]{0.35\textwidth}
		\includegraphics[width=1\columnwidth]{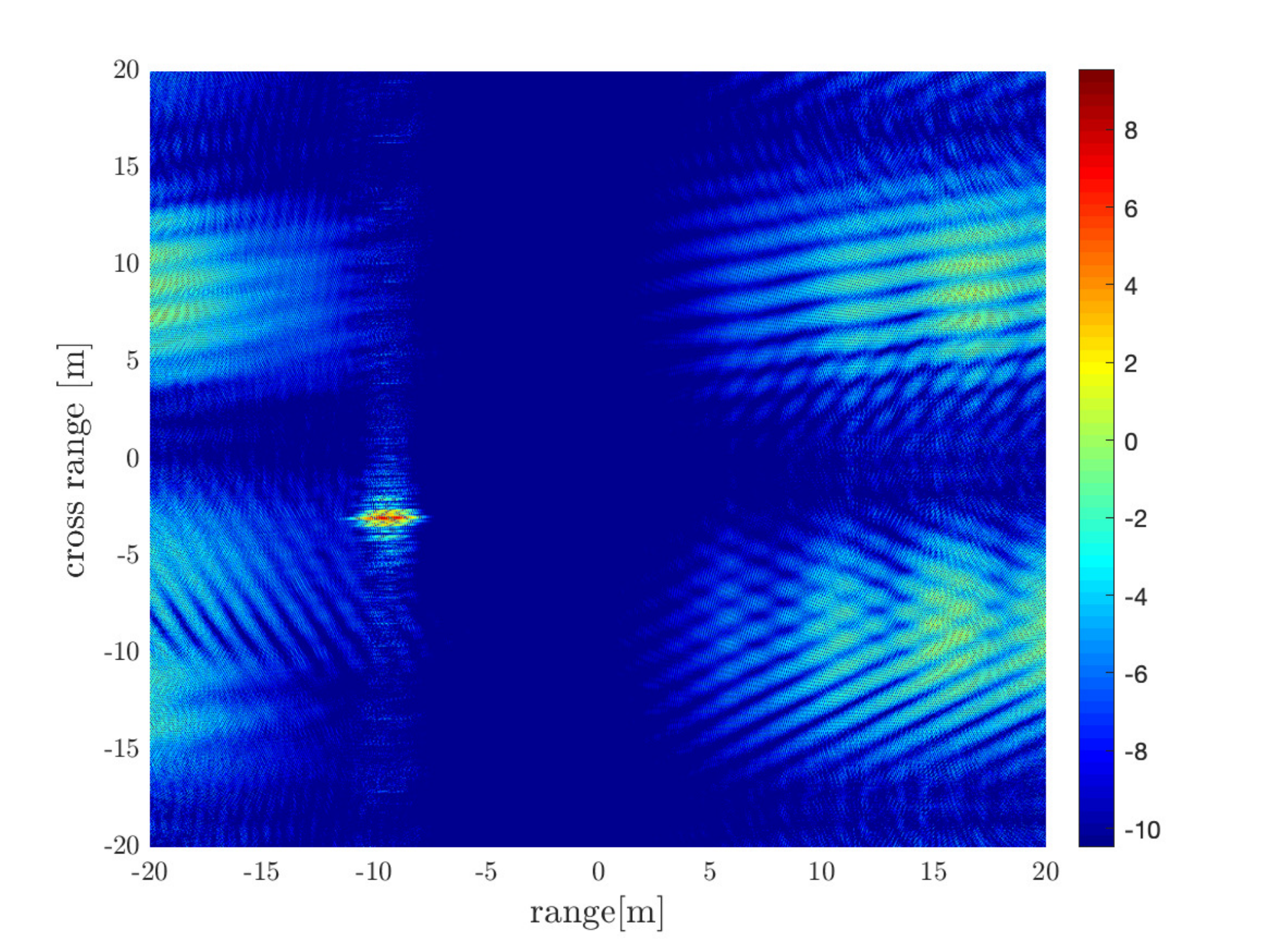}
		\caption{KM with exact motion estimation parameters applied to the sparse matrix generated by the conventional choice of $\eta$. }		
		\label{fig:image_S_orig}
	\end{subfigure}
	\begin{subfigure}[t]{0.35\textwidth}
		\includegraphics[width=1\columnwidth]{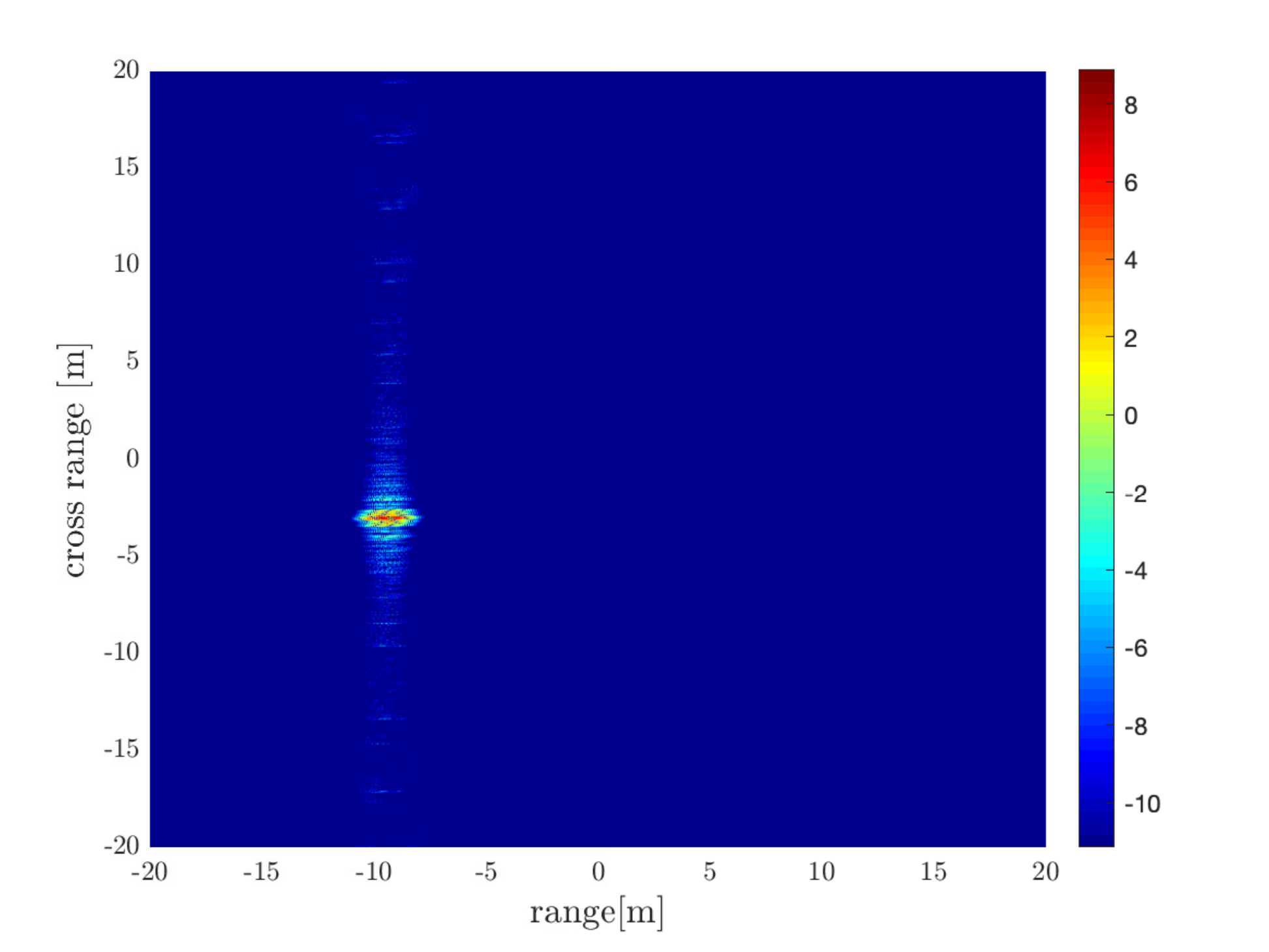}
		\caption{KM with exact motion estimation parameters applied to the sparse matrix generated by the optimal choice of $\eta$.}
		\label{fig:image_S_opt}
	\end{subfigure}
	\caption{Imaging applied to RPCA-separated data of Fig.~\ref{fig:RPCA}. We observe a significant improvement in the SNR when the optimal $\eta$ is used in the $\cL+\cS$  decomposition. }\label{fig:image}
\end{figure}

The corresponding Kirchhoff Migration (KM) images are shown in Figure \ref{fig:image}.  We observe that in the KM image of the original data matrix the moving target is not detectable (see Figure \ref{fig:image_D}). We also see that a lot of clutter is present in the KM image of the sparse matrix generated by the conventional choice of $\eta$ (see Figure \ref{fig:image_S_orig}). This is the result of stationary target's energy leakage into the sparse part. The KM image of the sparse matrix generated by the optimal choice of $\eta$ is shown in Figure \ref{fig:image_S_opt}. We observe a significant improvement in SNR compared to the results obtained with the conventional choice of $\eta$. As expected  improvement in the $\cL+\cS$ separation implies increased SNR for both the moving and the stationary targets' images.

\section{Analysis of RPCA for motion estimation}
\label{RPCA:anal}
We carry out in this section the analysis of the RPCA algorithm for motion estimation. This will allow us to determine the optimal range of values for the parameter~$\eta$.
\subsection{RPCA behavior as a function of $\eta$}
\label{sect:analysis}
The conventional value for the parameter $\eta$ is $\frac{1}{\sqrt{\max(n_1,n_2)}}$, where $n_1,n_2$ are the dimensions of the data matrix to be analyzed (cf. \cite{candesRPCA}). This choice is motivated by assuming that the sparse part is random noise, sparsely supported in the data matrix. However, this is not the case for the SAR data. 

By taking into account the specific form of $\cL$ and $\cS$ for the SAR problem, we derive an optimal range of values for $\eta$. Our analysis is based on the following observation: The ability of RPCA to separate between stationary and moving targets is due to the difference in their corresponding matrix norms.  The two necessary conditions for the algorithm to achieve separation are:
\begin{equation}
\begin{split}
\|A\|_*<\eta \|A\|_1,\hspace{1em} \text{for all low rank matrices } A,\end{split}
\end{equation}
and 
\begin{equation}
\begin{split}
\|A\|_*>\eta \|A\|_1,\hspace{1em} \text{for all sparse matrices } A.
\end{split}
\end{equation}
These conditions suggest the following restrictions on $\eta$ 
\begin{equation}
\begin{split}
&\eta > \frac{\|A\|_*}{\|A\|_1}, \hspace{1em}\text{ for all low rank matrices } A , \\
\quad&\eta < \frac{\|A\|_*}{\|A\|_1}, \hspace{1em}\text{ for all sparse matrices }A.
\end{split}
\end{equation}
We can therefore define the following range of values for $\eta,\quad \eta_{\min} \le \eta \le \eta_{\max}$, 
\begin{equation} \label{eq:etaminmax}
\begin{split}
 &\text{ with }\eta_{\min}= \sup\limits_{A \text{ low rank}}\frac{\|A\|_*}{\|A\|_1},\hspace{1em} \eta_{\max}=  \inf\limits_{A \text{ sparse}}\frac{\|A\|_*}{\|A\|_1}.
\end{split}
\end{equation}
Note that \eqref{eq:etaminmax} is a necessary condition but not sufficient. Thus, it provides only an estimate of the range of allowed values for $\eta$. An optimal $\eta$ can be determined by minimizing the objective
\begin{equation}
\begin{split}
	&\min\limits_{\eta}\hspace{0.5em}F(\eta)=\frac{\eta_{\min}}{\eta}+\frac{\eta}{\eta_{\max}}\\
	&\Rightarrow\eta^*=\sqrt{\eta_{\min}\eta_{\max}},\hspace{1em}F(\eta^*)=2\sqrt{\frac{\eta_{\min}}{\eta_{\max}}} .
	\end{split}
\end{equation}
We next analyze each of the terms, i.e., $\eta_{\min}$ and $\eta_{\max}$ to arrive at a closed form expression for $\eta^*$. We then show that using this $\eta^*$ does yield results that are more robust, meaning that there is no need for fine tuning the windowing of the SAR data and $\cL+\cS$ separation is achieved for a wider range of moving target's velocities. 

We also note that our analysis implies that $\eta$ is independent of any global matrix normalization, i.e. $\frac{\|\lambda A\|_*}{\|\lambda A\|_1}=\frac{\|A\|_*}{\|A\|_1}$, and, assuming that there is little variance in the norm of the rows of the SAR data matrix, $\eta$ is independent of any row normalization as well.

\subsection{Estimation of matrix norms in the SAR model}
\label{anal:origl1}
We choose as a representative of each class of matrices, i.e. sparse and low rank, a data matrix for a single target, either stationary or moving. We then compute the corresponding matrix norms in both cases and arrive at analytical expressions for $\eta_{\min}$ and $\eta_{\max}$ as in \eqref{eq:etaminmax}.  We also show with numerical simulations that these matrices which correspond to a single stationary and moving target are indeed good representatives of the classes of low rank and sparse matrices for SAR. 
Using this norm evaluations we get analytical expressions for $\eta_{\max}$ and $\eta_{\min}$ and hence we can compute the optimal value of $\eta$ for the SAR motion separation problem.
In the following sections we carry out evaluations of the matrix  $\ell_1$ and nuclear norms for stationary and moving targets. The main results are presented here and the reader is referred to \ref{app:B} for the detailed technical calculations.

\subsubsection{Estimation of the $\ell_1$ norm for the original SAR data matrix}
The $\ell_1$ norm is given by
\begin{equation}
\|\cD\|_1=\sum\limits_{i,j}|\cD_{ij}|.
\label{anal_l1}
\end{equation}
From here on, we assume for convenience a Gaussian baseband pulse 
\begin{equation}
\label{eq:gaussian}
f_B(t) = e^{-B^2t^2/2}
\end{equation}
and given our model form \eqref{eq:Dr_time} we have
\[ 
\begin{split}
&\sum\limits_{i,j}|\cD_{ij}|=\\
&\sum\limits_{i=1}^{n+1}\sum\limits_{j=1}^{m+1} |\cos(\om_o (t_j-\Delta\tau(s_i)) |e^{-B^2 (t_j-\Delta\tau(s_i))^2/2}.
\end{split}
\] 
The sum for every time trace can be well approximated as a Riemann integral,  
\begin{equation}
\sum\limits_{i,j}|\cD_{ij}|\approx\sum\limits_{i=1}^{n+1}\frac{1}{ \Delta t}\int\limits_{-\infty}^{\infty}|\cos(\om_ot)|e^{-B^2 t^2/2}dt.
\label{eq:L1_orig_1}
\end{equation}
We notice that if $\Delta\tau(s_i)\in [t_0,t_m], \forall i,\  \Delta t \ll \{1/B,\om_o\}$, then the absolute sum of elements is approximately the same for all time traces,
\begin{equation}
\sum\limits_{i,j}|\cD_{ij}|\approx\frac{2S(a)}{\Delta s \Delta t}\int\limits_{-\infty}^{\infty}|\cos(\om_ot)|e^{-B^2 t^2/2}dt.
\label{eq:L1_orig_2}
\end{equation}
Here $2S(a)=n\Delta s$, is the aperture size (in time units), along the trajectory.
In \ref{app:A} we provide an estimate of this integral using the Poisson summation formula. We get
\begin{equation}
\begin{split}
\sum\limits_{i,j}|\cD_{ij}|\approx \frac{2S(a)}{\Delta s B\Delta t} 2\sqrt{\frac{2}{\pi}}+O(e^{-2\omega_o^2/B^2}).
\end{split}
\label{eq:L1_orig_3}
\end{equation}
 Since $B\ll \om_o$, the approximation is highly accurate. We also notice that there is no difference in the value of the $\ell_1$ norm between stationary and moving targets. Thus, we infer that, in the SAR case, the main difference between stationary and moving targets is due to their corresponding nuclear norms.

\subsubsection{Estimation of the $\ell_1$ norm for the baseband SAR data matrix}

The $\ell_1$ norm is given by
\begin{equation}
\|\cDb\|_1=\sum\limits_{i,j}|\cDb\hspace{0.1mm}_{ij}|.
\end{equation}
Given our model form \eqref{eq:BB_form} and assuming a Gaussian pulse as in \eqref{eq:gaussian} we have
\begin{equation}
\sum\limits_{i,j}|\cDb\hspace{0.1mm}_{ij}|=\sum\limits_{i=1}^{n+1}\sum\limits_{j=1}^{m+1}  e^{-B^2 (t_j-\Delta\tau(s_i))^2/2}.
\end{equation}
Similar to \eqref{eq:L1_orig_1}, \eqref{eq:L1_orig_2}, we can approximate this sum using a Riemann integral as follows, 
\begin{equation}
\sum\limits_{i,j}|\cDb\hspace{0.1mm}_{ij}|\approx\frac{2S(a)}{\Delta s\Delta t}\int\limits_{-\infty}^{\infty}e^{-B^2 t^2/2}dt=\frac{2 S(a)}{\Delta s B\Delta t}\sqrt{2\pi}.
\label{eq:L1_BB_1}
\end{equation}
We notice again that there is no difference in the value of the $\ell_1$ norms between stationary and moving targets. Notice also that the $\ell_1$ norm is larger by a factor of $\pi/2$ compared to the original SAR data matrix (see \eqref{eq:L1_orig_3}).

\subsection{Spectral properties of the data matrix}

In order to evaluate the spectrum of the SAR data matrix we look at the eigenvalues of $\cD^T\cD$, whose $(j,k)-$th element is the inner product of the $j$-th and $k$-th columns of $\cD$.
By asymptotic analysis we see that for a single target the form of 
\begin{equation}
\cM=\cD^T \cD
\end{equation}
is of a block Gaussian matrix. Taking into account the effect of the target's speed qualitatively, we see that the magnitude of the diagonal elements of $\mathcal{M}$ will be inversely proportional to the slope of the target's trace, while the total block size of the matrix $\cM$ is proportional to the target's speed. This is true because the faster the target is moving, the larger the column support the time traces will have in the data matrix. Taking $B\Delta t\rightarrow\infty$, we can approximate $\cM$ by a diagonal matrix, whose size and elements are related by
\begin{equation}
\mathcal{M}^N\in\mathbb{R}^{N\times N}\subset \mathbb{R}^{m\times m},\quad \mathcal{M}^N_{ij}=\frac{\alpha}{N} \delta_{ij}.
\label{eq:diag_lim}
\end{equation}
Hence in this case the spectrum of $\cM$ is simply $\sqrt{\frac{\alpha}{N}}$, and the nuclear norm of $\cM$ is given  by 
\begin{equation}
\sum\limits_{i=1}^N \sqrt{\frac{\alpha}{N}}=\alpha \sqrt{N}
\end{equation}
i.e, the nuclear norm grows with the square root of $N$ and consequently with the square root of the target's speed. Notice that the $\ell_2$ norm of the singular values does not change and is always equal to $\alpha$. This is what is expected, because this is the Frobenius norm of $\mathcal{D}$, which is insensitive to translation in the fast time and doesn't discriminate between a moving and a stationary target. This preliminary analysis suggests that we should expect a growth rate of the nuclear norm with the target's speed of the form $cv_t^\beta$ with $\beta \lesssim 0.5$.

More detailed calculations given in \ref{app:B} show that, the spectrum has a Gaussian decay form as in 
\begin{equation}
\sigma_k^2\propto e^{-\frac{\pi^2k^2}{N^2 B^2\Delta t^2}},
\label{spec:gauss}
\end{equation}
where as mentioned already $N$, the column range of the target, is an increasing function of the velocity.


Because the sum of $\sigma_k^2$ is constrained by the Frobenius norm to be independent of $N$ and since 
$\sum\limits_{k=0}^N e^{-\frac{\pi^2k^2}{N^2 B^2\Delta t^2}}\propto N$,
we deduce that the proportionality factor is $\propto \frac{1}{N}$. Thus, the nuclear norm is proprtional to $\sqrt{N}$, as expected. As we shall see, it is this spectral property that makes RPCA a good tool for separating stationary from moving targets in SAR data. The nuclear norm is a sparsity-promoting norm in the spectrum. Thus, even though for our class of matrices the sum of the squares of singular values  is invariant with respect to the target's velocity, the nuclear norm depends on it. 

As pointed out in \ref{app:B}, the form of \eqref{spec:gauss} is based on a linear approximation of $\Delta\tau(s)$. For targets that are moving at low velocities, or at directions close to parallel to the direction connecting the imaging region and the platform, the spectrum deviates from this Gaussian form. However, it turns out that the nuclear norm is still well approximated by considering that the spectrum has this Gaussian decay form with respect to $N$. 
This seems to stem from the fact that the singular values always obey the constraint, $\sum\limits_k \sigma_k^2 =C$, almost independently of the target's velocity or location. Thus, the nuclear norm of two spectra will be similar as long as the two spectra have similar support, which explains the robustness of our approximation for the nuclear norm evaluation.

\subsection{Optimal Value for the RPCA parameter, $\eta^*$}
Based on the evaluations of the matrix norms, we obtain the optimal $\eta$, $\eta^*=\sqrt{\eta_{\min} \eta_{\max}}$ with
$$
\eta_{\min}=\sqrt{\frac{\Delta s B\Delta t}{4 S(a)\sqrt{\pi}}}   
$$
\[ 
\begin{split}
\eta_{\max}=\sqrt{\frac{\Delta s B\Delta t}{4 S(a)\sqrt{\pi}}}   \frac{ 1}{\sqrt{\frac{ N(v_t)B\Delta t}{2\sqrt{\pi}}+\frac{1}{2}}} \frac{\frac{\sqrt{2}N(v_t)B \Delta t}{\pi}+1}{2},
\end{split}
\] 
where, to leading order,
$$ 
N(v_t)\approx\frac{4S(a) }{\Delta t }\frac{1}{\|\vr(0)-\vrho_o(0)\|}(\vr(0)-\vrho_o(0))\cdot \frac{v_t}{c}
$$
is the minimal column support of the moving target in the data matrix, which is in turn determined by the minimal velocity to be resolved. For the derivation of these expressions see \ref{app:B}. 

Note that, as expected, the column support $N(v_t)$ grows with either a larger aperture or a smaller time resolution, as well as with the velocity component in the direction parallel to the vector connecting the platform to the imaging region. Thus, velocity resolution does not depend only on the speed but also on the direction in which the target is moving. Targets moving in a transverse direction would exhibit very small variation in their time delays independently of their speed.

We also see a growth of $\eta^*\propto (N(v_t) B\Delta t)^{1/4}$. Notice that as $N(v_t) B\Delta t\nearrow \infty$, we approach the diagonal limit of \eqref{eq:diag_lim}, which exhibits square root like growth of the nuclear norm with $N$. On the other hand, as $N(v_t) B\Delta t\searrow 0$, the dynamic range becomes smaller, and it is harder to separate the targets, as the time traces have wider and wider support. In that case, the variance of the traces with an increasing slope of $\Delta \tau(s)$ would be less significant. We conclude that both the target's trajectory/speed and the probing signal characteristics affect the performance of RPCA. 

\subsection{Baseband transformation and RPCA}
By qualitatively estimating the nuclear norm of matrices associated with both stationary and moving targets, we have determined how the spectrum and, consequently, the nuclear norm depends on the target's velocity. Using these evaluations, and the corresponding ones for the $\ell_1$ norms of the data matrices, we computed an optimal value for the parameter $\eta$ determined by the minimally resolved velocity.

\begin{figure}[htbp]
	\centering
	\begin{subfigure}[b]{0.32\textwidth}
	\includegraphics[width=1\columnwidth]{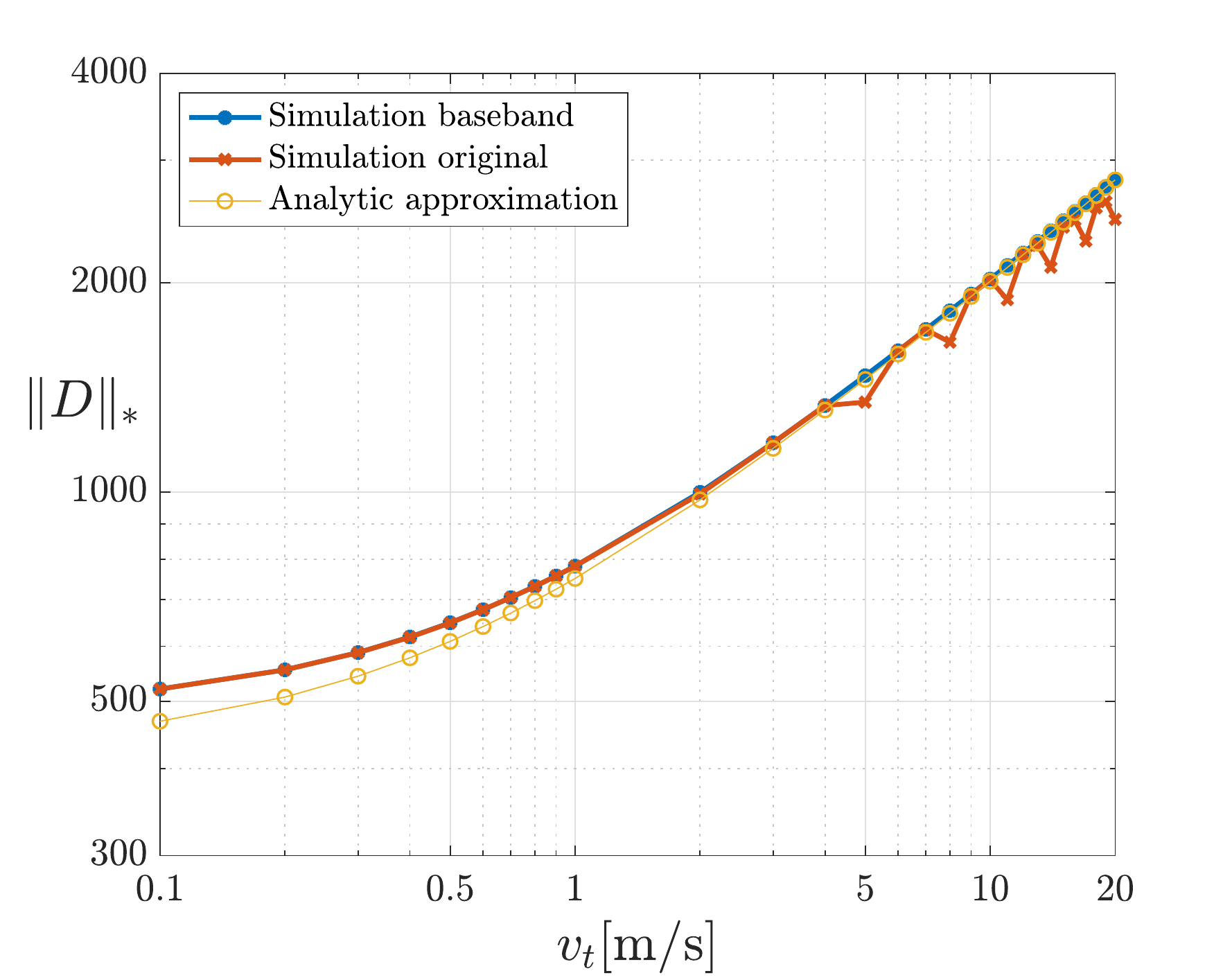}
	\caption{}
	\label{fig:nuc_norm_comp_a}
\end{subfigure}
\begin{subfigure}[b]{0.32\textwidth}
	\includegraphics[width=1\columnwidth]{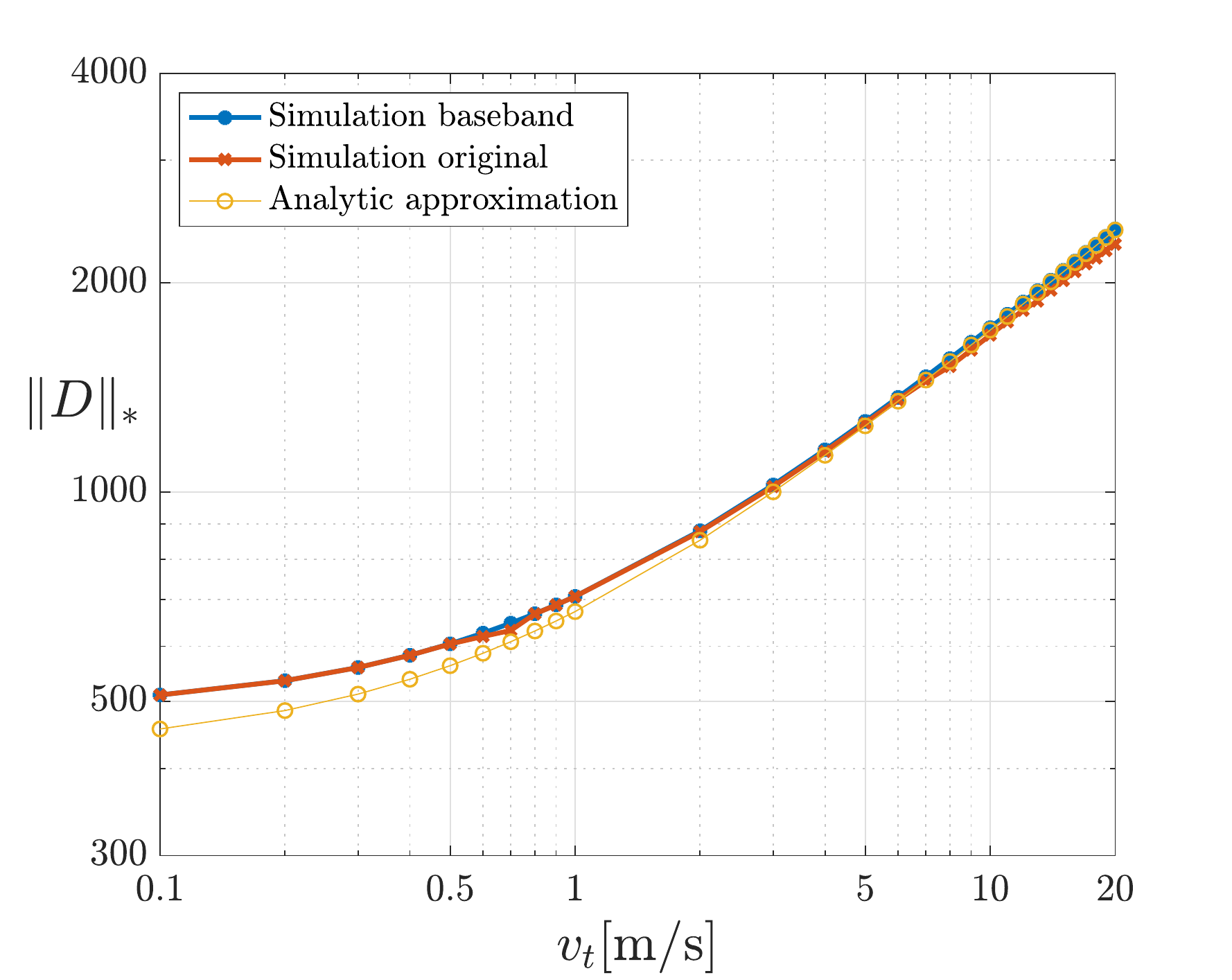}
	\caption{}
	\label{}
\end{subfigure}
\begin{subfigure}[b]{0.32\textwidth}
	\includegraphics[width=1\columnwidth]{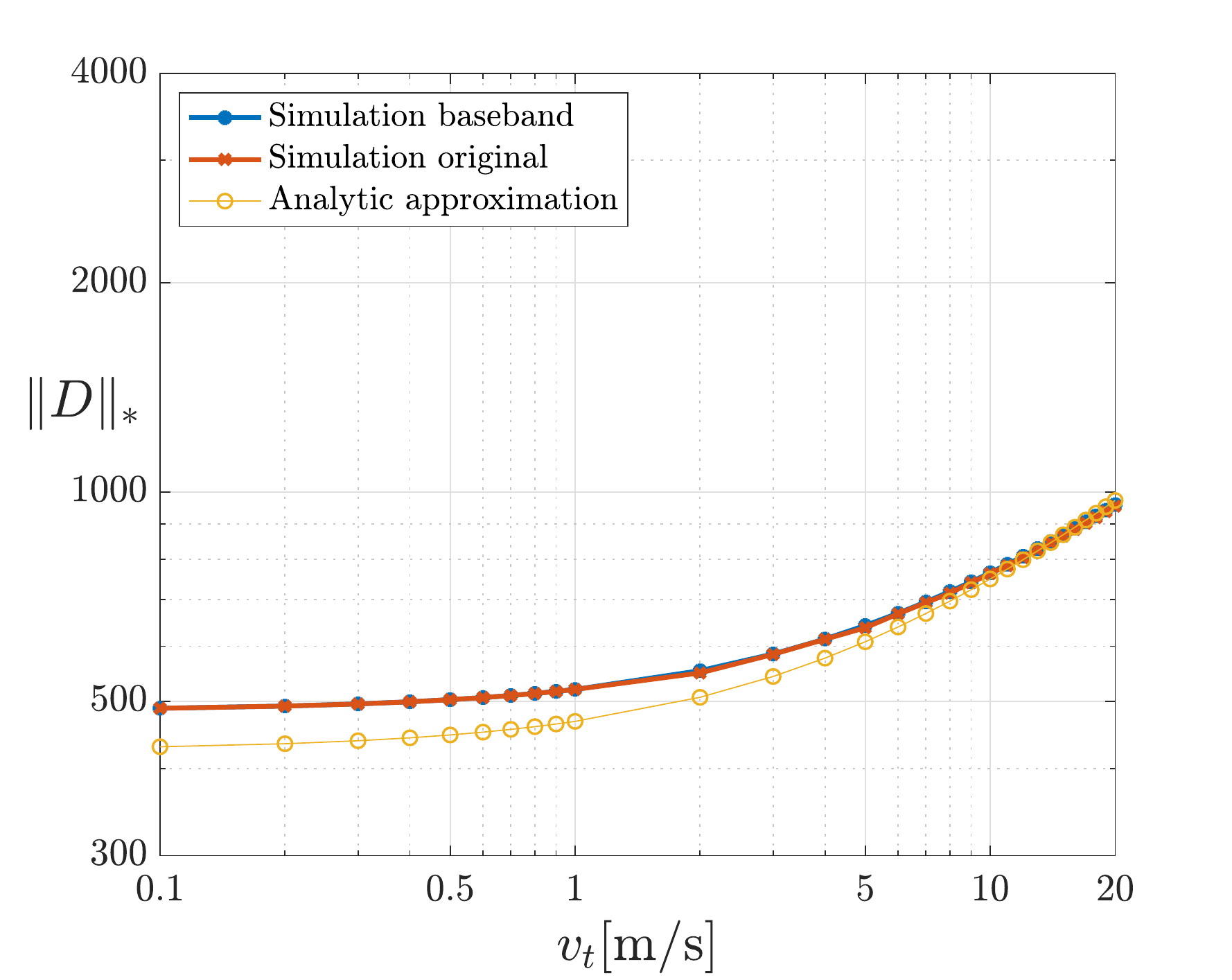}
	\caption{}
\end{subfigure}
\caption{Comparison of the analytic approximation of the spectrum for the original and babseband data matrices for a target at a direction with angle with respect to the $x$ axis of $(a)$ $\frac{\pi}{32}$,$(b)$ $\frac{\pi}{4}$,$(c)$ $\frac{15\pi}{32}$. We can see that the estimates are better for the baseband matrix, especially at smaller angles. See also Appendix \ref{app:B:2} for further results.}
\label{fig:nuc_norm_comp}
\end{figure}

Our asymptotic analysis results suggest that similar performance is expected for RPCA for both the original and the baseband data matrices. However, the asymptotic analysis, used to derive the optimal parameter, $\eta^*$, is highly simplified when using the baseband transformed matrix. Moreover, as shown in Figure \ref{fig:nuc_norm_comp}, the agreement between the asymptotic evaluation of the nuclear norm and the corresponding numerically obtained value is better in the baseband case.

\begin{figure}[htbp]
	\centering
	\begin{subfigure}[b]{0.24\textwidth}
		\includegraphics[width=1.05\columnwidth]{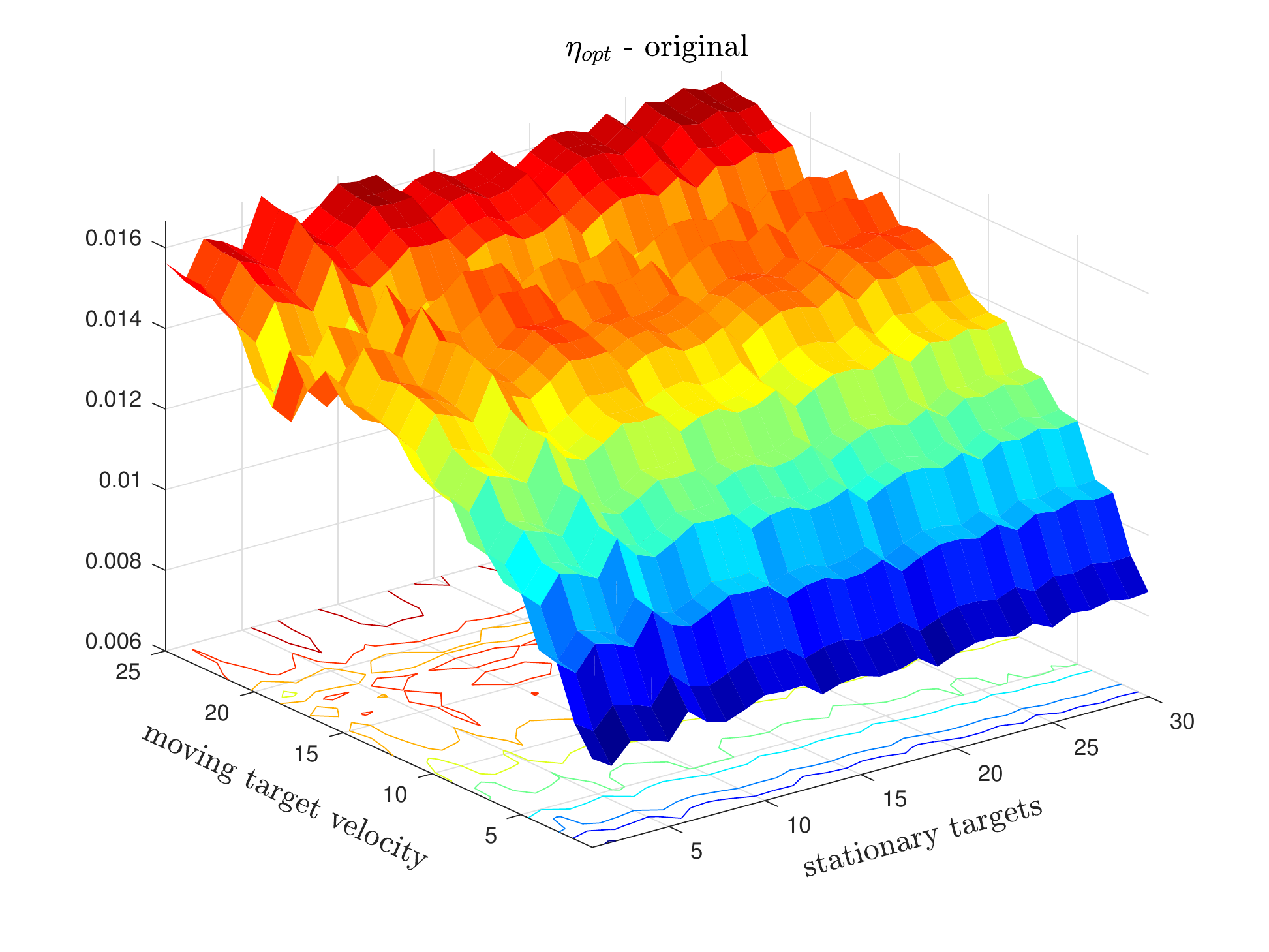}
		\caption{}
		\label{fig:eta_opt_orig}
	\end{subfigure}
	\begin{subfigure}[b]{0.24\textwidth}
		\includegraphics[width=1.05\columnwidth]{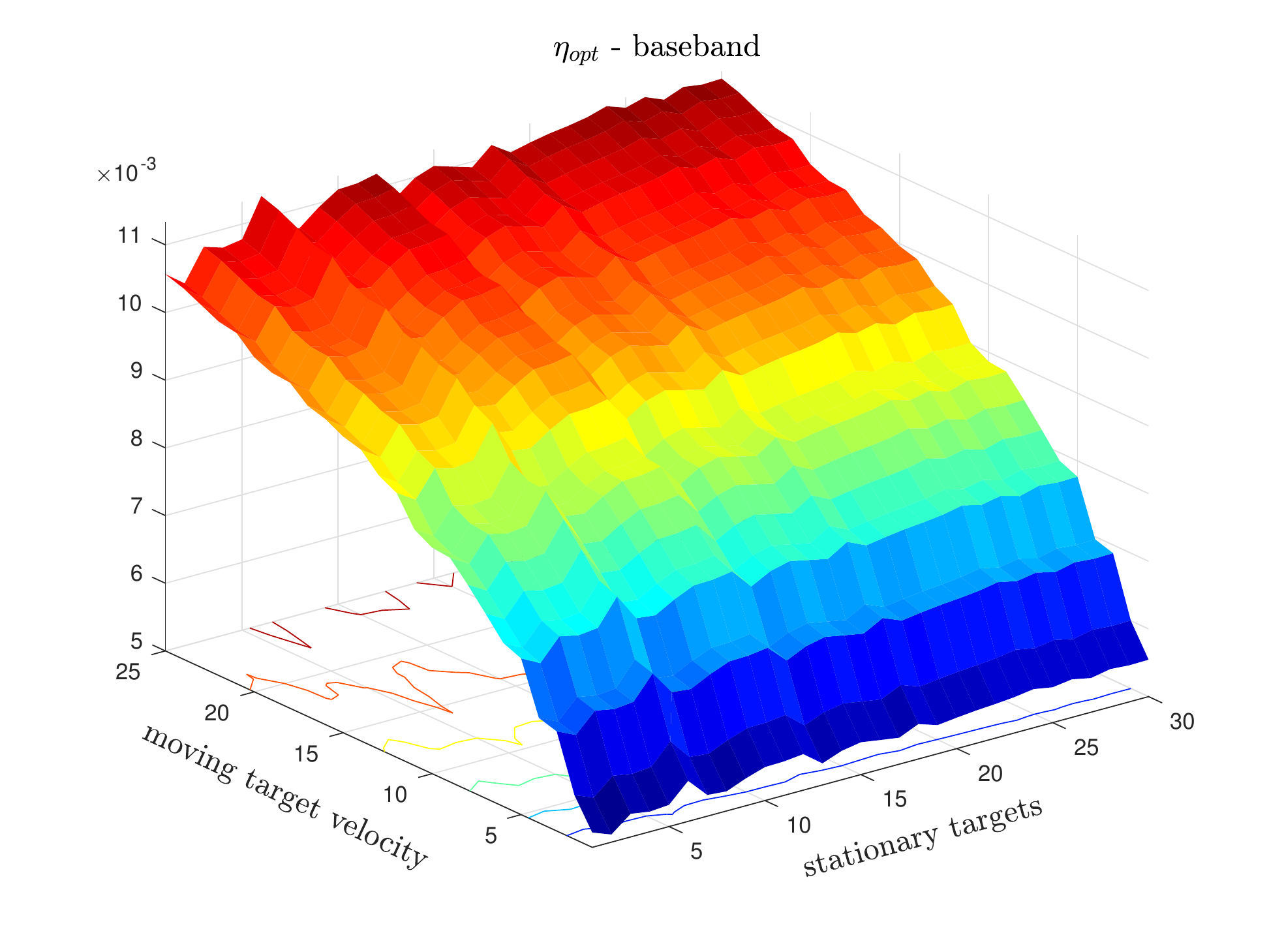}
		\caption{}
		\label{fig:eta_opt_BB}
	\end{subfigure}
		\begin{subfigure}[b]{0.24\textwidth}
		\includegraphics[width=1.05\columnwidth]{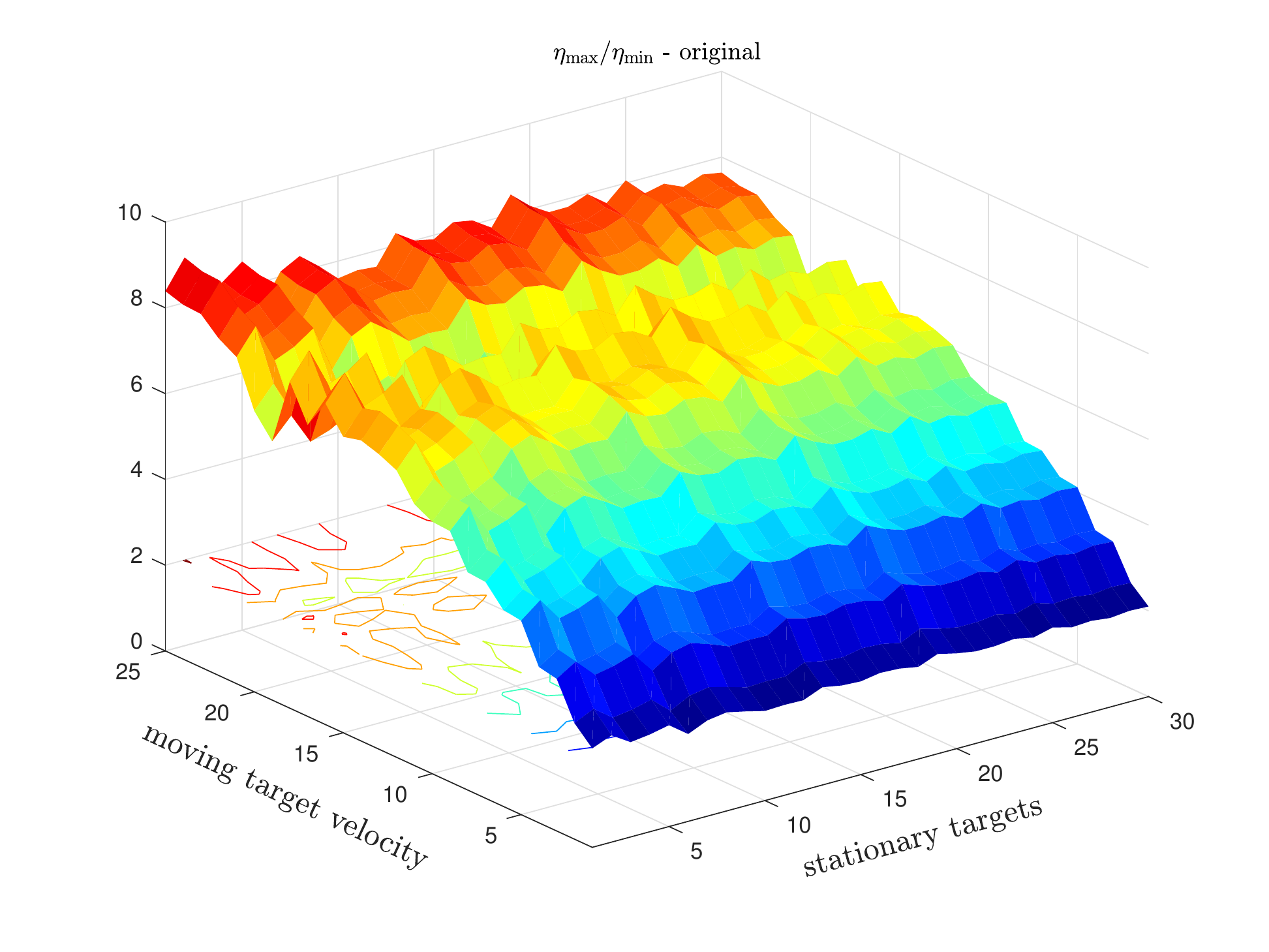}
		\caption{}
		\label{fig:eta_dyn_orig}
	\end{subfigure}
	\begin{subfigure}[b]{0.24\textwidth}
		\includegraphics[width=1.05\columnwidth]{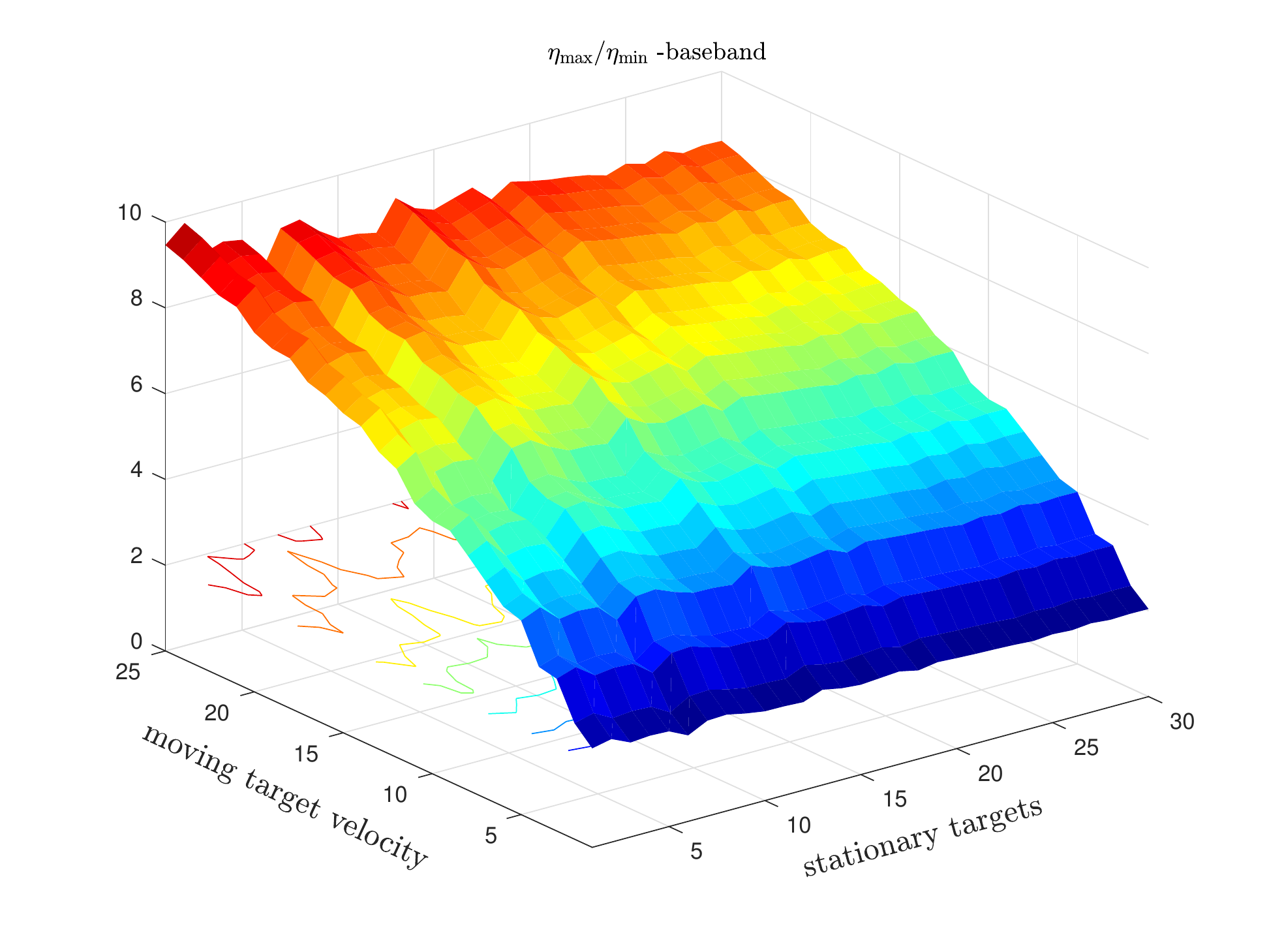}
		\caption{}
		\label{fig:eta_dyn_bb}
\end{subfigure}
\caption{Simulation results showing the optimal value of $\eta$. We simulated different cases evaluating $\eta_{\max},\eta_{\min}$ for different scenarios, varying the number of stationary targets and the velocity of the moving target. Top row: $\eta^*=\sqrt{\eta_{\max}\eta_{\min}}$. Bottom row $\sqrt{\frac{\eta_{\max}}{\eta_{\min}}}$, which is an indicator of the dynamic range of admissible values of~$\eta$.}
\label{fig:eta}
\end{figure}

Additionally, our numerical simulations provide evidence that the RPCA algorithm in the baseband case has an improved performance in terms of the dynamic range that $\eta$ can take as illustrated by Figure~\ref{fig:eta}.  We observe in general a much smoother behavior in the baseband case.  We also see very little variance of $\eta$ with changing the number of stationary targets compared with varying the velocity. This justifies our choice of considering a single stationary target as the representative of the class of stationary data matrices. We also see that the dynamic range of $\eta$ for the baseband matrices is larger, suggesting an overall better performance of the separation when applying RPCA to the baseband SAR data matrices.

\newpage
\section{Conclusions}
In this paper, we considered the synthetic aperture radar (SAR) imaging problem and focused our attention in the separation of the echoes coming from moving targets from those due to the stationary objects in the scene. Depending on the complexity of the scene, the velocity and the scattering properties of the moving target, this problem can be quite challenging. SAR imaging is designed to account only for stationary targets and the presence of moving objects introduces artefacts which may significantly affect the the image's SNR. Moreover, the moving target's echoes should not be considered as noise but as a useful signal that if processed adequately will reveal the velocity and position of the target. We addressed this problem using the robust principal component analysis that separates a matrix $\cD$ into its low rank, $\cL$, and sparse $\cS$ parts, i.e., $\cD =\cL+\cS$, by minimizing the objective functional  $\|\cL\|_*+ \eta \| \cS \|_1$. For the SAR problem, the low rank part consists of the stationary targets' echoes while the sparse part contains the echoes of the moving targets. Our main contribution is a detailed analysis that allows us to provide a range of optimal values for the single parameter $\eta$, which weights the $\ell_1$ norm of $\cS$ in the minimization functional. Our analysis suggests that the key quantity is the nuclear norm of the low rank part, $\| \cL\|_*$, which is shown to be proportional to the square root of the velocity of the target. More precisely, it is the projection of the target velocity along the direction that connects the SAR platform to the imaging region that matters. This is intuitively expected since targets that are moving slowly or  parallel to the SAR platform will appear as stationary.  Our analysis also shows that better separation results are obtained with increasing bandwidth and fast time sampling rate.    Another important remark is that it is beneficial to apply the RPCA after transforming the SAR data to baseband. This increases the robustness of the separation with respect to the synthetic aperture and the moving targets speed.  Our numerical simulations carried out in the GOTCHA regime for X-band SAR surveillance are in very good agreement with our theoretical results. 


\renewcommand{\thesection}{Appendix \Alph{section}}
\setcounter{section}{0}
\renewcommand{\thesubsection}{\Alph{section}.\arabic{subsection}}
\renewcommand{\thesubsubsection}{\Alph{section}.\arabic{subsection}.\arabic{subsubsection}}
\renewcommand{\theequation}{\Alph{section}.\arabic{equation}}
\section{Parameters used in numerical simulations}
Simulations were generated using a platform moving linearly  at $V_p=300$m/s in the $y$ direction, located at $\vr(0)=[7.1\text{km},0,7.3\text{km}]$ at $s=0$.  The emitted signal has a frequency $f_0=9.6$GHz, and a bandwith of 622MHz ($B=311$MHz).
The total aperture size is 237 pulses, with $\Delta s$=0.015s. 

For figures 1, and 2, the imaging area is centered around $\vrho_o=[0,0,0]$. The stationary targets are located at $[4.67,-4.35,0]$, $[2.06,9.61,0]$, 
$[-3.02,10.64,0]$, $[1.27,-11.1,0]$, $[-4.4,-7.81,0]$. The moving target is located at $[-9.43,-3.07,0]$ at $s=0$. All target locations are given in meters.

For figures 4,6 and 7: The moving target is located at $5.8,7.29,0]$ at $s=0$. 

For figure 5, the stationary targets location were drawn uniformly in a $10$m$\times 10$m imaging area. The moving target is moving at the $x$ direction. 

\section{Estimates of the $L^1$ norm of the SAR data matrix}
\label{app:A}
We wish to calculate $\sum\limits_{i,j}|\cD_{ij}|$, which is
$$
\sum\limits_{i,j}f_B(t_j-\Delta\tau(s_i))|\cos(\omega_o(t_j-\Delta\tau(s_i)))|.
$$
Assuming $\tau(s_i)$ is behaving regularly ($\Delta\tau(s_i)\in[t_{\min},t_{\max}]\forall i$), the sum over different time traces (rows) is the same and independent of the local translation $\Delta \tau(s_j)$,
$$
\sum\limits_{i,j}|\cD_{ij}|\approx \frac{4S(a)}{\Delta s\Delta t}\int\limits_{-\infty}^\infty e^{-B^2t^2/2}|\cos(\omega_o t) |dt.
$$ 
The integral can be recast as a series, $\sum\limits_{n=-\infty}^{\infty}g(n)$,
$$g(n)=\int\limits_0^{\frac{\pi}{2\omega_o}}\cos(\omega_o x)[e^{-B^2(x+\frac{n\pi}{\omega_o})^2} +e^{-B^2(\frac{(n+1)\pi}{\omega_o}-x)^2}]dx.
$$
Then, using Poisson's summation formula, we have 
$$\sum\limits_{n=-\infty}^{\infty}g(n)=\sum\limits _{p=-\infty}^{\infty}\hat{g}(p),\quad \hat{g}(p)=\int\limits_{-\infty}^\infty g(y)e^{-i2\pi p y}dy
$$
and
$$
\hat{g}(p)=2\sqrt{\frac{2}{\pi B^2}}(-1)^pe^{-2\omega_o^2p^2/B^2}.
$$
Thus,
$$\sum\limits_{p=-\infty}^{\infty}\hat{g}(p)\approx 2 \sqrt{\frac{2}{\pi B^2}}+O(e^{-2\omega_o^2/B^2}).
$$
Since in our setting $\omega_o\gg B$ we can approximate the integral well as $2 \sqrt{\frac{2}{\pi B^2}}$. Notice the $\frac{2}{\pi}$ factor with respect to the regular Gaussian integral $\sqrt{\frac{2\pi}{B^2}}$, which is the $\ell_1$ norm for the baseband data matrix. Therefore we deduce that the $\ell_1$ norm is smaller by a factor of $2/\pi$ compared to the baseband data case.
\section{Estimates of the spectrum of the SAR data matrix}
\label{app:B}
\subsection{Baseband case}
\label{app:B:1}
Given the data matrix $\Db(s,t)$, we want to evaluate its spectrum, based on the number of targets (stationary and moving) and the size of the aperture. Following the model presented in Section ~\ref{sec:BB_matrix}, we evaluate the matrix $\Db^H\Db$, where the general form of $\Db$ for a single target is
\begin{equation}
\label{eq:DB}
\Db(s_i,t_j)=e^{i\omega \Delta \tau(s_i)}e^{-\frac{B^2(t_j-\Delta \tau (s_i))^2}{2}},\\
\end{equation}
with 
$$\Delta\tau (s_i)=\frac{2}{c}\left(\|\vr(s_i)-\vrho_t(s_i)\|-\|\vr(s_i)-\vrho_o\|\right).$$
When evaluating the matrix $\Db^H\Db$, the complex phase of  \eqref{eq:BB_form} cancels out. We are left with
\begin{equation}
\begin{array}{l}
\Db^H\Db(j,k)\\
=\sum\limits_ie^{-\frac{B^2(t_j-\Delta \tau (s_i))^2}{2}}e^{-\frac{B^2(t_k-\Delta \tau (s_i))^2}{2}}\\
=e^{-\frac{B^2}{2}(t_j^2+t_k^2)}\sum\limits_ie^{-B^2(-(t_j+t_k)\Delta \tau(s_i)+(\Delta \tau(s_i))^2}\\
=e^{-\frac{B^2}{2}(t_j^2+t_k^2)}\sum\limits_ie^{-B^2\phi(s_i)}
\end{array} \label{eqbb2}
\end{equation}
%
%
with $\phi(s)$ defined as 
$$\quad \phi(s)=-(t_j+t_k)\Delta \tau(s)+(\Delta \tau(s))^2.$$
We can aproximate the last sum in \eqref{eqbb2} as an integral
$$
\sum\limits_i e^{-B^2\phi(s_i)}\approx\frac{1}{\Delta s}\int\limits_{s^-}^{s^+}e^{-B^2\phi(s)}ds.
$$

The values of $\Delta \tau$ range as the same value of $t$ (to be shown), thus $B^2\max|\Delta \tau|^2\gg1$, and we can evaluate the integral using Laplace's method
$$
\int\limits_{s^-}^{s^+}e^{-B^2 \phi(s)}ds\approx e^{-B^2\phi(s^*)}\sqrt{\frac{\pi}{B^2|\phi''(s^*)|}}.
$$
The equation for $s^*$ is
$$
\frac{d\phi(s)}{d s}\Big|_{s^*}=\left(-\frac{t_j+t_k}{2}+\Delta \tau(s)\right)\frac{d \Delta \tau(s)}{d s}\Big|_{s^*}=0.
$$
So, 
$$
\mbox{either } \Delta\tau(s^*)=\frac{t_j+t_k}{2} \mbox{ or } \Delta \tau'(s^*)=0.
$$
The first option is the only possibility if $\Delta \tau $ is close to linear. It would also dominate, as it achieves by definition the minimum value $g(x)=-(t_j+t_k)x+x^2$ and $\phi(s)=g(\Delta\tau(s))$.
Assuming $\exists s^*\in [s^-,s^+]$ we have $\Delta\tau(s^*)=\frac{t_j+t_k}{2}$, and
$$\phi(s^*)=-\frac{(t_j+t_k)^2}{4},\quad \phi''(s^*)=2\Delta \tau'(s^*)^2.
$$
Hence,
\begin{equation}
\begin{split}
\label{eq:DBTDB}
\Db^H\Db(j,k)&\approx e^{-\frac{B^2}{2}(t_j^2+t_k^2)}e^{-B^2\phi(s^*)}\sqrt{\frac{\pi}{B^2|\phi''(s^*)|}}\\
&=e^{-\frac{B^2}{4}(t_j-t_k)^2}\sqrt{\frac{\pi}{B^2|\phi''(s^*)|}}\\
&=e^{-\frac{B^2}{4}(t_j-t_k)^2}{\frac{\sqrt{\pi}}{B|\Delta\tau'(s^*)|}}.
\end{split}
\end{equation}


Note that for $\Delta \tau(s)$ quasi-linear, $\Delta\tau'(s)$ is close to a constant, and the matrix has a T\"{o}plitz form. If no such stationary point exists, the element would be strongly suppressed. So we expect, for a single stationary target, $ \Db^H\Db$ to have the form of a matrix, with a localized block around the diagonal. Notice $\Delta \tau(s^*)=\left(\frac{t_j+t_k}{2}\right)$, so the matrix $\Db^H\Db(j,k)$ has the approximate general form
	$$
	C_{j+k}e^{-\frac{B^2}{4}(t_j-t_k)^2}\mathbbm{1}_{\{\Delta \tau_{\min}\le t_j,t_k\le \Delta \tau_{\max}\}}
$$
i.e, a T\"{o}plitz times Hankel form on a subset of indices of size $N=\frac{ \Delta \tau_{\max}- \Delta \tau_{\min}}{\Delta t}$. $\Delta \tau_{\min},\Delta\tau_{\max}$ are the extreme values of $\Delta\tau(s), s\in[s^-,s^+]$.

We can approximate 
\begin{equation}
\begin{split}
&\Delta \tau(s)= \frac{2}{c}\left(\|\vr(s)-\vrho_t(s)\|-\|\vr(s)-\vrho_o\|\right)\\
&\approx \frac{2}{c}\left[ -\frac{(\vr(s)-\vrho_o)\cdot(\vrho_t(s)-\vrho_o))}{\|\vr(s)-\vrho_o\|}+\frac{1}{2}\frac{\|\vrho_t(s)-\vrho_o\|^2}{\|\vr(s)-\vrho_o\|}\right].
\end{split}
\label{spec:del_tau}
\end{equation}
If $\Delta \tau(s)$ is approximately linear, then $\phi''(s)$ is independent of $s$, $C_{j+k}\equiv C$, and the matrix has a T\"{o}plitz form. 

The dominant $s$ dependent term is $\frac{2}{c}\left[ -\frac{(\vr(0)-\vrho_o)\cdot v_t s}{\|\vr(s)-\vrho_o\|}\right]$, thus the analysis holds as long the target's motion has a significant portion perpendicular to the platforms movement.

	\subsubsection{Evaluation of the matrix's spectrum}
	Using results from \cite{gray2006toeplitz}, we can approximate the spectrum for large $N$ using a circulant matrix. The first row of the circulant matrix would be
	$$
	\mathcal{C}_j= C e^{-\frac{B^2}{4}t_j^2}+C  e^{-\frac{B^2}{4}(N\Delta t-t_j)^2},\quad j=0,...,N-1.
	$$
	
	The spectrum of $\mathcal{C}$ is given by the Discrete Fourier transform of $\mathcal{C}_{1,j}$, which we again approximate for large $N$ as an integral 
	\[ 
	\begin{split}
	\lambda^{\mathcal{C}}_k&=\sum\limits_{j=0}^{N-1}\mathcal{C}_{0,j}e^{-i\frac{2\pi j k} {N}}\\
	&=\sum\limits_{j=0}^{N-1} C  e^{-\frac{B^2}{4}t_j^2}e^{-i\omega_k t_j }+Ce^{-\frac{B^2}{4}(N\Delta t-t_j)^2}e^{-i\omega_kt_j }\\
	&=\sum\limits_{j=-N+1}^{N-1}C  e^{-\frac{B^2}{4}t_j^2}e^{-i\omega_kt_j }
	\approx \int\limits_{-\infty}^{\infty}e^{-\frac{B^2}{4}t^2}e^{-i\omega_k t} \\
	&\propto e^{-\frac{\omega_k^2}{B^2}},\quad \omega_k = \frac{2\pi k}{N\Delta t}, k=0,...,N-1.
	\end{split}
	\] 

	When evaluating the integral we need to take $\omega_k\rightarrow \omega_k \mod [-\pi,\pi]$, thus the right end of the spectrum actually corresponds to negative values of $\omega$. This would give us the following spectrum (in decreasing value)
	$$\lambda_k^{\mathcal{C}}=\sigma_k^2\approx C e^{-\frac{\pi^2k^2}{N^2 B^2\Delta t^2}}.
	$$
	Using the fact that $\sum\limits_k \sigma_k^2=\|\Db\|_F^2$, and we have similar calculation to \eqref{eq:L1_BB_1} $\|\Db\|_F^2=\frac{2S(a)\sqrt{\pi}}{\Delta s B\Delta t}$
	$$
	\sum\limits_{k=0}^{N-1} \sigma_k^2=\frac{2S(a)\sqrt{\pi}}{\Delta s B\Delta t}.
	$$
	Formally
	$$
	\sum\limits_{k=0}^{\infty} e^{-\frac{\pi^2k^2}{N^2 B^2\Delta t^2}}=\frac{1}{2}\vartheta_3(0,e^{-\frac{\pi^2}{N^2 B^2\Delta t^2}})+\frac{1}{2}
	$$
	where $\vartheta_3$ is the Jacobi theta function of the third kind. $\vartheta_3(0,e^{-\frac{1}{\sigma^2}})$ can be well approximated as
	\begin{equation}
	\frac{1}{2}\vartheta_3(0,e^{-\frac{\pi^2}{N^2 B^2\Delta t^2}})+\frac{1}{2}=\begin{cases} 1, & \sigma\lesssim \frac{1}{\sqrt{\pi}},\\ \frac{\sqrt{\pi}\sigma+1}{2},&\sigma\gtrsim \frac{1}{\sqrt{\pi}}. \end{cases}
	\label{jac_sum}
	\end{equation}
Thus, for large $N$ and $B\Delta t\ll \pi$, we can approximate the sum using the Jacobi theta function as
	\begin{equation}
	\sum\limits_{k=0}^{N-1} \sigma_k^2\approx\alpha \left(\frac{N B\Delta t}{2\sqrt{\pi}}+\frac{1}{2}\right),
	\end{equation}
	i.e.,
	$\alpha\approx\frac{ 2S(a)\sqrt{\pi}}{ \Delta s B\Delta t \left(\frac{NB\Delta t}{2\sqrt{\pi}}+\frac{1}{2}\right)}
$. We can compute $N(v_t)=\frac{ \Delta \tau_{\max}- \Delta \tau_{\min}}{\Delta t}$ using our data model. 

We can approximate $N(v_t)$, for targets that have a non negligible component of their velocity in the direction of $\vr-\vrho_o$  using \eqref{spec:del_tau} as
\begin{equation}
N(v_t)\approx \frac{4S(a)}{c\Delta t}\frac{1}{\|\vr(0)-\vrho_o(0)\|}(\vr(0)-\vrho_o(0))\cdot v_t \\
\label{spec:N_v}
\end{equation}
%
If $(\vr(0)-\vrho_o(0))\cdot v_t$ gets too small compared to higher order terms, then the linear behavior is no longer accurate, but using \eqref{spec:del_tau}, we get an accurate quadratic approximation.

Thus, the expression for the spectrum would be
$$
\sigma_k\approx\
\sqrt{\frac{ 2S(a)\sqrt{\pi}}{ \Delta s B\Delta t\left(\frac{N B\Delta t}{2\sqrt{\pi}}+\frac{1}{2}\right)} } e^{-\frac{\pi^2k^2}{2N^2 B^2\Delta t^2}}, 
$$
and therefore, 
\[ 
\begin{split}
&\|\Db\|_*\approx
\sqrt{\frac{S(a)\sqrt{\pi}}{\Delta s B\Delta t}}\max\left(1,\frac{ \frac{\sqrt{2}NB \Delta t}{\pi}+1}{2\sqrt{\frac{ N B\Delta t}{2\sqrt{\pi}}+\frac{1}{2}}}\right).
\end{split}
\] 
Since $N\propto |v_t|$ we can expect a $\propto \sqrt{v_t}$ growth in the nuclear norm with increasing velocity.

For the approximation of \eqref{jac_sum} to be valid we need
\begin{equation}
\frac{N B\Delta t}{\pi}\gtrsim \frac{1}{\sqrt{\pi}}\Rightarrow N\gtrsim \frac{\sqrt{\pi}}{B\Delta t}.
\end{equation}
This means that for smaller $N$ we use explicit summation to normalize the spectrum.  Figure
\ref{fig:nuc_norm_comp} shows numerical simulations that suggest that this approximation of the nuclear norm holds well even when the matrix deviates from the T\"{o}plitz form. Thus our estimate of the spectrum is robust for a broad range of target directions.
%
%
%
%
%

\subsection{Spectrum of the original data matrix}
\label{app:B:2}
We can repeat the analysis of the previous section for the original data matrix.
Givent that the data model for a single target is
\[ 
\begin{split}
&D(s_i,t_j)=\cos(\omega_o(t_j- \Delta \tau(s_i))e^{-\frac{B^2(t_j-\Delta \tau (s_i))^2}{2}}.
\end{split}
\] 

Evaluating the matrix, we have the expression
\[ 
\begin{split}
&(D^HD)_{jk}=\frac{1}{2}\cos(\omega_o(t_j- t_k))e^{-\frac{B^2}{2}(t_j^2+t_k^2)}\times \\
&\hspace{1em}\big(\sum\limits_ie^{-B^2(-(t_j+t_k)\Delta \tau(s_i)+(\Delta \tau(s_i))^2}\big)
\\&\hspace{0.25em}+\frac{1}{2}e^{-\frac{B^2}{2}(t_j^2+t_k^2)}\big(\sum\limits_i \cos(\omega_o(t_j+t_k-2\Delta\tau(s_i)))\times\\
&\hspace{1em}e^{-B^2(-(t_j+t_k)\Delta \tau(s_i)+(\Delta \tau(s_i)))^2}\big).
\end{split}
\] 
Notice the first term behaves exactly like the term in the baseband case. We approximate the second  term again as an integral and obtain 
\[ 
\begin{split}
&\approx\frac{1}{4}e^{-\frac{B^2}{2}(t_j^2+t_k^2)}\frac{1}{2\Delta s}\int\limits_{s^-}^{s^+}e^{-g_+(s)}+e^{-g_-(s)}ds\\
&g_\pm(s)=\pm i\omega_o(t_j+t_k-2\Delta\tau(s_i))\\
&-B^2(-(t_j+t_k)\Delta \tau(s_i)+(\Delta \tau(s_i))^2).
\end{split}
\] 
Stationary phase analysis gives the dominant stationary point
$s^*_\pm=\frac{t_j+t_k}{2}\mp i\frac{\omega_o}{B^2}.$

\begin{figure}[htbp]
	\centering
		\includegraphics[width=0.31\textwidth]{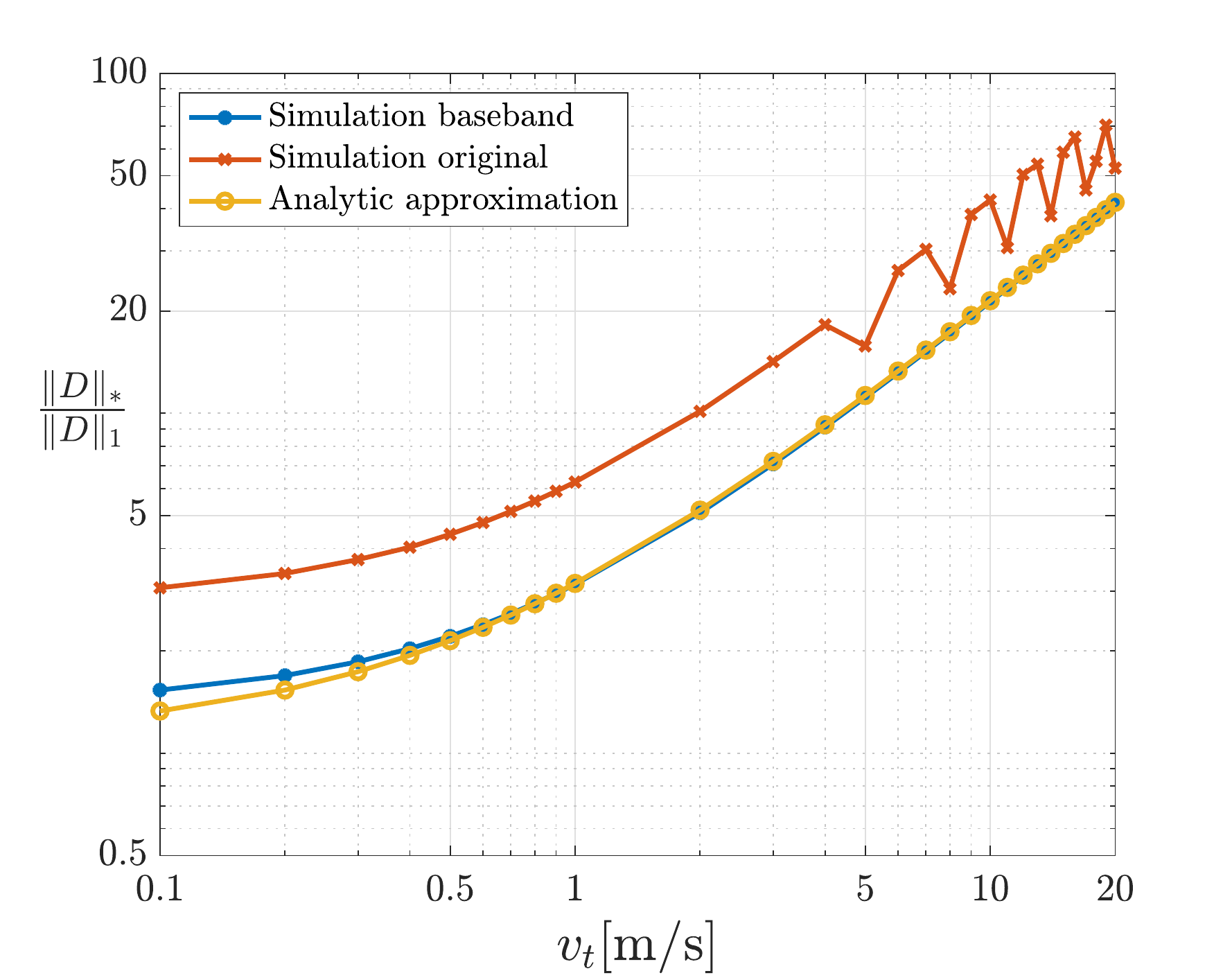}
		\caption{$\|D\|_*/\|D\|_2$;  
		same parameters as in Fig~\ref{fig:nuc_norm_comp_a}. 
		The evaluation is closer to the analytical model for the baseband data, though the difference is smaller when looking at the nuclear norm (see Fig.~\ref{fig:nuc_norm_comp_a}). }

		\label{fig:spec_comp}
\end{figure}

Plugging this back into the expression we get:
$$
\frac{1}{4}e^{-\frac{B^2}{2}(t_j^2+t_k^2)}e^{-g_\pm(s^*_\pm)}=\frac{1}{4}e^{-\frac{B^2}{4}(t_j-t_k)^2}e^{-3\frac{\omega_o^2}{B^2}}.
$$
Since $\omega_o\gg B$ this term is heavily attenuated, so that we can neglect it, resulting to
	\[ 
	\begin{split}
&D^HD(j,k)\approx \frac{1}{2}\cos(\omega_o(t_j- t_k))Ce^{-\frac{B^2}{4}(t_j-t_k)^2}
\end{split}
\] 
for $ t_j,t_k\in[\Delta \tau_{\min},\Delta \tau_{\max}]$, and 0 otherwise.
We next approximate $D^HD$ as a circulant matrix, and use the DFT to calculate its spectrum, denoted as $\mu_k^2$. The effect of the cosine factor would be to create two copies of the spectrum of the baseband matrix, $\sigma_k^2$, each multiplied by a factor of $\frac{1}{4}$. Thus, the spectrum is 
\begin{equation}
\mu_k^2= \frac{1}{4}\sigma_{\lfloor k/2\rfloor}^2.
\end{equation}
Note that this is consistent with
\begin{equation}
\|D\|_F^2=\sum\limits_k\mu_k^2=\frac{1}{2}\sum\limits_k\sigma_k^2=\frac{1}{2}\|\Db\|_F^2.
\end{equation}
Which is the expected result, from the element-wise Frobenius norm computation. This would also mean 
\begin{equation}
\|D\|_*=\|\Db\|_*.
\end{equation}
As shown in Figures~\ref{fig:nuc_norm_comp} and \ref{fig:spec_comp}, simulations suggest that the actual evaluation of the nuclear norm is more accurate for the baseband case over a wider range of parameters.

\section*{Acknowledgments}
The work of M. Leibovich and G. Papanicolaou was partially supported by AFOSR FA9550-18-1-0519. The work of C. Tsogka was partially supported by AFOSR FA9550-17-1-0238 and AFOSR FA9550-18-1-0519. 
\bibliographystyle{plain} \bibliography{SBIR}

\end{document}